\documentclass[fleqn,usenatbib]{mnras}

\usepackage{graphicx} 
\usepackage{amsmath}  
\usepackage{amssymb}  

\usepackage{booktabs}
\usepackage{xspace}
\usepackage{acronym}
\usepackage[tight]{units}
\usepackage[english,capitalize]{cleveref}
\graphicspath{{./figures/}}

\newcommand{\LSm}{LS220$^\dagger$\@\xspace}
\newcommand{\LSl}{LS175$^\dagger$\@\xspace}
\newcommand{\LSh}{LS255$^\dagger$\@\xspace}
\newcommand{\msm}{$m^*_{0.8}$\@\xspace}
\newcommand{\msS}{$m^*_{\textnormal{S}}$\@\xspace}
\newcommand{\msKS}{$(m^* K)_{\textnormal{S}}$\@\xspace}
\newcommand{\msKES}{$(m^* K E_{\textnormal{sym}})_{\textnormal{S}}$\@\xspace}

\newcommand{\del}[2]{\frac{\partial #1}{\partial #2}}
\newcommand{\dens}[1]{\unit[10^{#1}]{g\,cm^{-3}}}
\newcommand{\ye}[0]{Y_{\textnormal{e}}}

\newcommand{\ie}{i.e.\@\xspace}
\newcommand{\eg}{e.g.\@\xspace}
\newcommand{\ceg}[1]{\cite[see, e.g.,][]{#1}}

\newacro{CFL}{Courant–Friedrich–Lewy}
\newacro{BNS}{binary neutron star}
\newacro{EOS}{equation of state}
\newacro{NS}{neutron star}
\newacro{BH}{black hole}
\newacro{GW}{gravitational wave}

\usepackage[T1]{fontenc}
\usepackage{ae,aecompl}

\usepackage{newtxtext,newtxmath}

\title[Nuclear matter properties in BNS mergers]
      {Effects of nuclear matter properties in neutron star mergers}

\author[M. Jacobi]{
    M. Jacobi$^{1}$\thanks{E-mail: mjacobi@theorie.ikp.physik.tu-darmstadt.de},
    F. M. Guercilena$^{2,3}$,
    S. Huth$^{1,4}$,
    G. Ricigliano$^{1}$,
    A. Arcones$^{1,5}$\thanks{E-mail: almudena.arcones@physik.tu-darmstadt.de},
    and A. Schwenk$^{1,4,6}$
    \\
    $^{1}$Institut für Kernphysik, Technische Universität Darmstadt, Schlossgartenstr. 2, Darmstadt 64289, Germany\\
    $^{2}$INFN-TIFPA, Trento Institute for Fundamental Physics and Applications, via Sommarive 14, I-38123 Trento, Italy\\
    $^{3}$Dipartimento di Fisica, Universit\`{a} di Trento, Via Sommarive 14, 38123 Trento, Italy\\
    $^{4}$ExtreMe Matter Institute EMMI, GSI Helmholtzzentrum für Schwerionenforschung GmbH, 64291 Darmstadt, Germany\\
    $^{5}$GSI Helmholtzzentrum für Schwerionenforschung GmbH, Planckstr. 1, Darmstadt 64291, Germany\\
    $^{6}$Max-Planck-Institut für Kernphysik, Saupfercheckweg 1, 69117 Heidelberg, Germany
}

\pubyear{2023}

\begin{document}
\label{firstpage}
\pagerange{\pageref{firstpage}--\pageref{lastpage}}
\maketitle


\begin{abstract}
  The dynamics in mergers of \ac{BNS} systems depend sensitively on the \ac{EOS} of dense matter.
  This has profound implications on the emission of \acp{GW} and the ejection of matter in the merger and post-merger phases and is thus of high interest for multi-messenger astronomy.
  Today, a variety of nuclear \acp{EOS} are available with various underlying microphysical models.
  This calls for a study to focus on EOS effects from different physical nuclear matter properties and their influence on \ac{BNS} mergers.
  We perform simulations of equal-mass \ac{BNS} mergers with a set of 9 different \acp{EOS} based on Skyrme density functionals.
  In the models, we systematically vary the effective nucleon mass, incompressibility, and symmetry energy at saturation density.
  This allows us to investigate the influence of specific nuclear matter properties on the dynamics of \ac{BNS} mergers.
  We analyze the impact of these properties on the merger dynamics, the fate of the remnant, disk formation, ejection of matter, and gravitational wave emission.
  Our results indicate that some aspects of the merger, such as the frequencies of the post-merger gravitational-wave spectrum and the shock-heated ejecta mass, are sensitive to the EOS around saturation density while others, such as the contraction of the remnant and the tidal ejecta mass, are sensitive to the behavior towards higher densities, e.g., characterized by the slope of the pressure versus density.
  The detailed density dependence of the \ac{EOS} thus needs to be taken into account to describe its influence on \ac{BNS} mergers.
\end{abstract}


\acresetall

\section{Introduction}
\label{sec:introduction}

The observation of the \ac{GW} signal GW170817 \citep{Abbott2017a, Abbott2018a, Abbott2019a} together with the corresponding short $\gamma$-ray burst GRB170817A and UV/optical/near-IR transient \citep{Abbott2017b} was the first detection of a \ac{NS} merger and of gravitational and electromagnetic waves from the same source.
Despite being the only event of its kind so far, it provided useful information for nuclear physics and astrophysics.
Several constraints to the nuclear \ac{EOS} such as the inference of the \ac{NS} tidal deformability from the measurement of the inspiral \ac{GW} signal \ceg{Abbott2017a, De2018a} or constraints on the maximum mass of a cold \ac{NS} \ceg{Margalit2017a, Rezzolla2018a, Shibata2019a, Most2020a, Nathanail2021a} could be deduced.
Furthermore, the observation of the kilonova AT2017gfo \citep{Andreoni2017a, Arcavi2017a, Coulter2017a, Cowperthwaite2017a, Diaz2017a, Drout2017a, Evans2017a, Hu2017a, Valenti2017a, Kasliwal2017a, Lipunov2017a, Pian2017a, Pozanenko2018a, Smartt2017a, Tanvir2017a, Troja2017a, Utsumi2017a} matched predictions for emission powered by radioactive decay of r-process elements \citep{Li1998a, Kulkarni2005a,   Metzger2010a, Kasen2013a, Tanaka2013a}.
This confirmed merging \ac{BNS} as an astrophysical site for the r-process \ceg{Rosswog2018a, Kawaguchi2018a} which was first theorized by \citet{Lattimer1974a}.
Still, whether compact object mergers are the dominant source of r-process in the Universe is up for debate \ceg{Cote2019a, Molero2021a}.

The nuclear \ac{EOS} is a key ingredient in \ac{BNS} merger simulations.
Experiments \ceg{Danielewicz2002a, Tsang2012a, Lattimer2013a, LeFevre2016a, Roca-Maza2015a, Russotto2016a, Birkhan2017a, Adhikari2021a, Adhikari2022a} and theoretical calculations based on chiral effective field theory \ceg{Hebeler2013a,Tews2013a,Lynn2016a, Drischler2019a, Drischler2020a, Keller2023a} constrain the \ac{EOS} up to around nuclear saturation density $n_0 = 0.16$\,fm$^{-3}$.
However, for densities inside \acp{NS} (two times $n_0$ or higher) the \ac{EOS} represents one of the largest sources of uncertainties in numerical models of \ac{BNS} mergers.
Observations of \acp{NS} can provide excellent constraints of cold matter at supra-nuclear densities.
The discovery of two-solar-mass neutron stars \citep{Demorest2010a, Antoniadis2013a, Fonseca2021a} has significantly reduced the uncertainties in the neutron star mass-radius relation.
These measurements suggest that in order to support neutron stars with such high mass, the \ac{EOS} cannot be too soft.
Furthermore, observations from NASA's NICER mission can constrain the mass and radius of pulsars simultaneously \citep{Riley2019a, Miller2019a, Riley2021a, Miller2021a} leading to novel constraints on the EOS \ceg{Raaijmakers2019a, Dietrich2020a, Legred2021a, Raaijmakers2021a, Huth2022a}.
Moreover, multi-messenger observations of binary compact object mergers like GW170817 / AT2017gfo can be used to probe the \ac{EOS} at finite temperatures and even higher densities.
However, to extract constraints from the observations of the merger of compact objects, detailed numerical relativity hydrodynamics simulations exploring the influence of many different \acp{EOS} are necessary.

The \ac{EOS} directly influences the dynamics of the merger and determines the fate and structure of the remnant and the \ac{GW} emission in the post-merger phase.
Shock heating and tidal torques can eject up to $10^{-2} M_\odot$ of material on dynamical timescales ($\unit[1-10]{ms}$) \citep{Rosswog1999a, Korobkin2012a, Hotokezaka2013a, Rosswog2013a, Sekiguchi2015a, Sekiguchi2016a}.
The masses of the very neutron-rich tidal ejecta and moderately neutron-rich shock-heated ejecta depend sensitively on the \ac{EOS}.
After the coalescence, a large fraction of the material ejected from the \acp{NS} stays gravitationally bound and forms an accretion disk around the central object.
The density in the disk is typically between $10^{10}- \dens{13}$ so in a regime where the \ac{EOS} is better constrained.
However, the disk mass, structure, and composition are determined by the merger dynamics.
Furthermore, as long as the remnant does not collapse to a \ac{BH} it interacts with the disk.
Therefore, the dynamics and mass ejection in the disk phase are indirectly influenced by the nuclear \ac{EOS}.
In the early disk phase ($\sim \unit[100]{ms}$ after the merger), mass is ejected by oscillations in the deformed remnant \citep{Bauswein2013b, Radice2018a, Nedora2019a, Nedora2021b}.
Furthermore, neutrinos emitted from the hot disk remnant interface can be reabsorbed in the outer layers of the disk supporting mass ejection \citep{Perego2014a}.
After $\sim \unit[1]{s}$ the disk cools down to a point where further cooling by neutrino emission becomes inefficient.
As a result, 10\% -- 50\% of the disk is ejected due to heating by an effective viscosity originating from the magnetorotational instability (MRI) \ceg{Fernandez2013a, Just2015a, Fujibayashi2018a, Fujibayashi2020a, Fahlman2022a, Hayashi2022a}.
The ratio of dynamical ejecta to disk outflow mass depends mainly on the fate of the central object.
If the remnant \ac{NS} does not collapse to a \ac{BH} the mass of the disk ejecta is one to two orders of magnitude larger than that of the dynamical ejecta \citep{Fujibayashi2020a}.
However, if a \ac{BH} is formed within $\unit[10-20]{ms}$ after the merger the ratio is closer to one \citep{Fujibayashi2023a}.

Many aspects of the nuclear \ac{EOS} influence the merger dynamics such as the detailed density dependence of the pressure, thermal effects, and the dependence on the composition.
Many studies aim to quantify nuclear physics uncertainties in numerical simulations by employing a small sample of different \ac{EOS} models \cite{Bovard2017b, Sekiguchi2016a, Radice2018a, Nedora2021a}, thereby varying many of the above-mentioned aspects at the same time.
While this approach is useful to reveal correlations with the general stiffness of the \ac{EOS}, the impact of its specific features cannot be studied this way.
Only a few works have focused on the dependence of \ac{BNS} mergers on individual aspects of the \ac{EOS}.
For example, \citet{Most2021a} investigated the impact of the slope of the symmetry energy on the post-merger dynamics and the temperature dependence has been investigated by changing the thermal pressure independently from the cold \ac{EOS} in several works \ceg{Bauswein2010a, Hotokezaka2013a, Raithel2021a, Fields2023a}.

In this work, we perform 3D general-relativistic simulations of merging \acp{NS} based on various \acp{EOS} with different nuclear matter properties.
The employed EOS models are based on the work of \citet{Yasin2020a} using Skyrme energy-density functionals combined with the liquid-drop model and are created with the SROEOS code \citep{Schneider2017a, Schneider2018a}.
By adjusting the parameters of the Skyrme functional, each nuclear matter property can be varied independently.
A similar approach was used in numerical simulations of supernovae \citep{Schneider2019a, Yasin2020a, Andersen2021a} and very recently also in \ac{BNS}-merger simulations \citep{Fields2023a}.
This enables us to study the impact of the nucleon effective mass, the incompressibility, the symmetry energy, and the saturation density and energy separately.
We specifically focus on the effective mass and incompressibility in this work.
Furthermore, we systematically vary all the nuclear matter properties mentioned from the values of the LS220 \citep{Lattimer1991a} to the ones of the Shen \ac{EOS} \citep{Shen1998a} as in the work of \citet{Yasin2020a}.

The paper is organized as follows. \Cref{sec:simulation_setup} describes our setup for performing \ac{BNS} merger simulations and explains the post-processing techniques used to analyze them. In \cref{sec:eos_tables}, we introduce the method to create the \ac{EOS} models as well as their key properties.
The results of the simulations are reported in \cref{sec:remnant,sec:gravitational_wave_emission,sec:ejecta_properties}, which describe the dynamics of the merger and post-merger phases, the post-merger gravitational wave emission, and the properties of the ejected material, respectively.
Finally, we summarize our conclusions in \cref{sec:conclusions}.


\section{Simulation setup}
\label{sec:simulation_setup}

The simulations we employ to model \ac{BNS} systems have been carried out employing the framework of the \texttt{EinsteinToolkit} suite \citep{Haas2020a, Loffler2012a}, itself based on the \texttt{Cactus} computational toolkit \citep{Goodale2003a}.

To handle general relativistic hydrodynamics we employ the \texttt{WhiskyTHC} code \citep{Radice2012a,Radice2014a,Radice2014b}.
Modeling the \ac{NS} matter as a perfect fluid, it solves Euler's equations for the balance of energy and momentum, coupled to conservation laws for the neutron and proton densities:
\begin{align}
	\nabla_\mu\left(n_{\textnormal{p,n}} u^\mu\right) &= R_{\textnormal{p,n}}\label{eq:continuity_euler1}\\
	\nabla_\nu T^{\mu\nu} &= Q u^{\mu}\,. \label{eq:continuity_euler2}
\end{align}
In \cref{eq:continuity_euler1,eq:continuity_euler2}, $n_{\textnormal{p,n}}$ are respectively the proton and neutron number densities, $u^\mu$ is the fluid four-velocity and $T^{\mu\nu}$ is the fluid stress-energy tensor.
In this formulation, due to charge neutrality, the electron fraction $\ye$ is directly related to the proton number density, \ie, $\ye \equiv n_{\textnormal{e}}/(n_p+n_n) = n_{\textnormal{p}}/(n_p+n_n)$.

\Cref{eq:continuity_euler1,eq:continuity_euler2} include source terms due to neutrino interactions: $R_{\textnormal{p,n}}$ are the net lepton number exchange rates, due to the absorption and emission of electron neutrinos and antineutrinos, while $Q$ is the net energy deposition rate due to the absorption and emission of (anti-) neutrinos of all flavors.
Neutrino emission and absorption are modeled with a leakage \citep{Galeazzi2013a, Neilsen2014a} scheme and the so-called ``M0'' scheme of \citet{Radice2018a}, respectively.
These are both ``grey'' (\ie, energy integrated) schemes evolving three neutrino species: electron neutrinos $\nu_{\textnormal{e}}$, electron antineutrinos $\bar{\nu}_{\textnormal{e}}$ and heavy neutrinos $\nu_{\textnormal{x}}$, which accounts for all others (anti-) neutrino flavors.
Furthermore, the M0 scheme evolves the distribution function of neutrinos on a ray-by-ray grid, which we set up consisting of 2048 rays uniformly spaced in latitude and longitude with a radial resolution $\Delta r\approx\unit[244]{m}$.

In our setup, \texttt{WhiskyTHC} employs a finite-volume scheme for the discretization of the hydrodynamic quantities.
The scheme reconstructs primitive variables with the fifth-order MP5 method \citep{Suresh1997a}, from which numerical fluxes are computed with the HLLE flux formula \citep{Harten1983a}, augmented with a positivity-preserving flux limiter \citep{Hu2013a,Radice2014a} in order to handle the transition to vacuum regions (which we fill with an atmosphere of density $\rho_{\textnormal{atmo}}\approx 1.24 \times \dens{3}$).

The hydrodynamics is coupled to a dynamically evolved spacetime.
Einstein's equations are written in the \mbox{BSSNOK} formulation \citep{Shibata1995a, Baumgarte1998a, Brown2009a}, and discretized with fourth-order finite-differences stencils by the \texttt{McLachlan} code \citep{Brown2009b}.
We furthermore choose the ``1+log'' and ``Gamma-driver'' gauge conditions \ceg{Baumgarte2021a}.

The time evolution is performed by the strong-stability-preserving RK3 integrator \citep{Gottlieb1998a} using a method-of-lines scheme.
The time step is determined by the \ac{CFL} criterion taking the speed of light as the maximum propagation speed.
The \ac{CFL} factor is chosen to be 0.15.

The mesh for our simulations is handled by the \texttt{Carpet} code \citep{Schnetter2004a}, which employs a ``moving boxes'' Berger-Oliger adaptive mesh refinement scheme \citep{Berger1984a, Berger1989a}.
We employ a Cartesian-coordinates computational domain made of 7 refinement levels.
The resolution on the finest level is $\unit[0.128]{M_\odot} \approx \unit[189]{m}$ and the full grid spans $\unit[1024]{M_\odot} \approx \unit[1512]{km}$ in each direction.
To reduce the computational cost, we only evolve the $z\ge0$ part of the domain, and impose reflecting boundary conditions at $z=0$.

\subsection{Postprocessing methods}
\label{sec:postprocessing}

We define the \ac{NS} as the region where the rest mass density is larger than $\dens{13}$.
Since the gauge is evolved dynamically during the simulation we need to compare simulation data in a gauge-independent way.
For this, we divide all grid cells in a snapshot of a simulation into 100 uniformly spaced density bins up to $\dens{15}$.
Any quantity of interest $A$ is then averaged in each bin, weighted by the conserved rest mass density $\sqrt{\gamma} \rho W$:
\begin{equation}
  \label{eq:rho_hist}
  \left| A \right|_\rho = \frac{\int_{\rho_{\text{bin}}} A \sqrt{\gamma} \rho W \,\mathrm{d}^3x}
  {\int_{\rho_{\text{bin}}} \sqrt{\gamma} \rho W  \,\mathrm{d}^3x}\,,
\end{equation}
where $\rho_{\text{bin}}$ is the rest mass density of the bin and $\int_{\rho_{\text{bin}}}\,\mathrm{d}^3x$ indicates the volume integral over all cells in the density bin.

Furthermore, we make use of the complex azimuthal mode decomposition given by \citep{Paschalidis2015a, East2016a, East2016b, Radice2016a, Nedora2021b}
\begin{equation}
  \label{eq:decomp}
  C_m = \int e^{- i m \phi} \sqrt{\gamma} W \rho \, \mathrm{d}x \mathrm{d}y \,,
\end{equation}
to study the deformation of the remnant.

We estimate the relevance of neutrino heating based on the net timescale of the neutrino heating $\tau_\nu$.
It is given by the ratio of the conserved internal energy density $\tau = E - D$ and the local net neutrino heating rate $Q_\nu = Q_\nu^{\text{M0}} - Q_\nu^{\text{Leak}}$:
\begin{equation}
  \label{eq:tau_nu}
  \tau_\nu = \frac{1}{\alpha W}  \frac{E-D}{Q_\nu^{\text{M0}} - Q_\nu^{\text{Leak}}}\,,
\end{equation}
where $\alpha$ is the lapse function.

The properties of the ejecta are extracted on a detection sphere with a radius of \unit[300]{km}.
The Bernoulli criterion is used to determine whether a fluid element is unbound.
It is defined by $-h u_t > h_\infty$, where $h_\infty = \lim_{\rho, T \rightarrow 0} h$ is the asymptotic specific enthalpy \citet{Foucart2021a}.
Accordingly, the asymptotic velocity of an ejected fluid element is defined as $v_\infty = \sqrt{ 1 - \left(\frac{h_\infty}{h u_t}\right)^2}$.
This criterion is less restrictive than the geodesic criterion $- u_t > 1$, which does not take the ejecta's thermal and binding energy into account.
Typically, the ejection rate of matter meeting the geodesic criterion stops after $\sim$ \unit[10]{ms} post-merger and roughly corresponds to the fluid elements ejected by dynamical processes (see \cref{sec:ejecta_properties} for further discussion).
Therefore, we define fluid elements fulfilling the geodesic criterion as dynamical ejecta \cite[the same convention is used by ][]{Nedora2021b, Combi2023a}.
Ejecta that only satisfy the Bernoulli, but not the geodesic criterion are defined as disk ejecta.
  The extraction radius of \unit[300]{km} ensures that the remnant accretion-disk system is fully contained (the disk typically extends to $\sim \unit[100]{km}$) but is small enough for different bursts of ejecta to be resolved.
  If instead a much larger extraction radius is chosen, the time at which ejected material is registered depends more on the ejecta velocity than on the time of ejection, which hinders the identification of different ejecta components.
For an in-depth discussion of different ejection criteria and their impact see, \eg, \citet{Foucart2021a}.

To extract \ac{GW} waveforms and spectra we employ the Newman-Penrose scalar $\Psi_4$ \citep{Newman1962a}, following Sect. 5 of \citet{Hinder2013a} for its practical implementation.
The second time derivative of the \ac{GW} strain polarization components $\ddot{h}_+$ and $\ddot{h}_\times$ can be related to $\Psi_4$ by:
\begin{equation}
  \label{eq:h_psi4}
  \ddot{h}_+ - i \ddot{h}_\times = \Psi_4 = \sum_{l=2}^\infty \sum_{m=-l}^l
  C_{lm}\,_{-2}Y_{lm}(\theta, \phi) \,,
\end{equation}
where $\Psi_4$ is expanded in weighted spherical harmonics of spin weight -2, $_{-2}Y_{lm}(\theta, \phi)$.
The expansion coefficients $C_{lm}$ are extracted at multiple finite coordinate radii inside the simulation domain and extrapolated to null infinity along outgoing radial null geodesics.
From them, the strain components $h_+$ and $h_\times$ are obtained by performing the time integration with the ``fixed frequency integration'' (FFI) method \citep{Reisswig2011a}.
Finally, the power spectral density (PSD) of the signal is given by
\begin{equation}
  \label{eq:PSD}
  \tilde{h}(f) = \sqrt{\frac{\left| \tilde{h}_+ (f) \right|^2
      + \left| \tilde{h}_\times (f) \right|^2}{2}} \,,
\end{equation}
with the frequency-domain strain components
\begin{equation}
  \label{eq:htilde}
  \tilde{h}_{+,\times}(f) = \int_0^\infty h_{+,\times}(t) e^{-i 2 \pi f t} \,\mathrm{d}t\,.
\end{equation}

\subsection{Overview of the BNS models}
\label{sec:overview}

We perform simulations for 9 \acp{EOS} in \cref{tab:eos_table} (see \cref{sec:eos_tables} for more details).
Each model initially consists of two irrotational identical $M=1.365 M_\odot$ \acp{NS} on quasi-circular orbits with an initial separation of \unit[45]{km}.
This combination corresponds to a chirp mass of $1.188 M_\odot$ and is thus compatible with the \ac{GW} source of GW170817.
This orbital separation corresponds to an inspiral phase of $2-3$ orbits before the merger.
The initial data for all the selected simulations are constructed using the spectral elliptic solver \texttt{LORENE} \citep{Gourgoulhon2001a}.
In the construction of the initial data, the minimum temperature slice of the \ac{EOS} table is used and the composition is determined by neutrinoless beta-equilibrium.

\begin{table*}
  \centering
  \caption{%
      Overview of all EOS models and their key results, including
    the simulated time $t_{\text{end}}$,
    time until collapse to a \ac{BH} $t_{\text{BH}}$,
    disk mass at the end of the simulation $M_{\text{disk}}$,
    masses of different ejecta components $M^{\text{ej}}$,
    frequencies of the two most prominent peaks in the post-merger \ac{GW} spectrum $f_1$ and $f_2$,
    radius $R_{\text{NS}}$ and reduced tidal deformability $\tilde{\Lambda}$ of the \acp{NS} in the initial data of our simulations. The EOS models and labels are based on \citet{Yasin2020a}.
    \label{tab:overview}}
  \begin{tabular}{lcccccccccccc}
    \toprule
    EOS & $t_{\text{BH}}$ & $t_{\text{end}}$ & $M_{\text{disk}}$ & $M^{\text{ej}}_{\text{tot}}$ &  $M^{\text{ej}}_{\text{tidal}}$ &
    $M^{\text{ej}}_{\text{shock}}$ & $M^{\text{ej}}_{\text{disk}}$ & $M^{\text{ej}}_{v_\infty>0.6c}$ & $f_1$ & $f_2$ & $R_{\text{NS}}$ & $\tilde{\Lambda}$ \\
    \footnotesize & [ms] &  [ms] & [$M_\odot$] & [$10^{-3} M_\odot$] & [$10^{-3} M_\odot$] & [$10^{-3} M_\odot$] & [$10^{-3} M_\odot$] & [$10^{-6} M_\odot$] & [kHz] & [kHz] & [km] \\
    \midrule
    \LSl   & 0.46 &   0.7 &     - &     - &     - &     - &     - &    -  &     - &     - & 12.16 &  361.0 \\
    \LSm   & 8.03 &  33.1 &  0.05 & 7.096 & 0.403 & 1.592 & 5.101 & 0.665 &  2.14 &  3.08 & 12.69 &  606.2 \\
    \LSh   &    - &  42.9 &  0.23 & 5.307 & 0.104 & 1.603 & 3.601 & 4.597 &  2.01 &  2.64 & 12.98 &  660.7 \\
    \msm   &    - &  44.9 &  0.14 & 7.717 & 0.138 & 1.111 & 6.469 & 2.050 &  1.87 &  2.82 & 12.95 &  699.6 \\
    \msS   &    - &  47.0 &  0.23 & 9.372 & 0.184 & 0.913 & 8.275 & 0.362 &  1.90 &  2.53 & 13.23 &  759.3 \\
    \msKS  &    - &  49.1 &  0.26 & 8.645 & 0.430 & 0.930 & 7.285 & 0.385 &  1.79 &  2.41 & 13.52 &  982.6 \\
    \msKES &    - &  47.5 &  0.23 & 5.887 & 0.021 & 0.883 & 4.982 & 0.467 &  1.72 &  2.30 & 14.06 & 1090.6 \\
    SkShen &    - &  54.1 &  0.23 & 7.939 & 0.004 & 0.550 & 7.384 & 0.079 &  1.66 &  2.25 & 14.49 & 1295.4 \\
    Shen   &    - &  55.3 &  0.24 & 9.527 & 0.001 & 0.645 & 8.882 & 0.066 &  1.68 &  2.24 & 14.55 & 1221.1 \\
    \bottomrule
\end{tabular}
\end{table*}

An overview of the simulations is compiled in \cref{tab:overview}.
Each simulation is run for at least \unit[40]{ms} post-merger.
The only exceptions are the models \LSl and \LSm.
In \LSl, a \ac{BH} is formed almost immediately after the merger so barely any mass is ejected.
In \LSm, the remnant collapses after $\sim$ \unit[8]{ms} which is enough time to eject matter and form an accretion disk.
However, the mass ejection and \ac{GW} emission stop completely after \unit[30]{ms}.
In all other models, the remnant is a massive \ac{NS} which stays stable for the duration of the simulation and is surrounded by an accretion disk.


\section{EOS models}
\label{sec:eos_tables}

In \ac{BNS} merger simulations, the influence of the \ac{EOS} on the dynamics of the merger is often discussed only in terms of the softness or stiffness of the \ac{EOS}, as encapsulated, e.g., in the radius of a cold non-rotating \ac{NS} or the reduced tidal deformability of the initial binary system \ceg{Bauswein2016a, Takami2015a, Rezzolla2016a, Nedora2021c}.
This approach has been successful in describing the peak frequencies of the post-merger \ac{GW} emission, typically within a $10\%$ error, as well as the threshold mass to prompt collapse \ceg{Bauswein2021a}.
The same approach has been extended to features of the remnant-disk system and the dynamical ejecta \citep{Dietrich2017b, Kruger2020a, Nedora2021b, Henkel2023a}.
In this work, in order to enable a more detailed description of the \ac{EOS}' features, we employ the expansion parameters of the energy density of nuclear matter at nuclear saturation density.

The energy per particle of nuclear matter $\frac{E}{A}$ at zero temperature is a function of number density $n=n_n + n_p$ and isospin asymmetry $\beta = \frac{n_n - n_p}{n} = 1 - 2 x$.
Here, $n_n$ and $n_p$ are the neutron and proton number densities, respectively, and $x=n_p/n$ is the proton fraction, which is equal to the electron fraction $\ye$ due to charge neutrality.
The energy per particle can be expanded around saturation density $n=n_0$ and symmetric matter $\beta=0$:
\begin{align}
  \left. \frac{E}{A} \right|_{T=0}  = \left. \frac{\epsilon}{n} \right|_{T=0}  \approx
  -B + \frac{1}{2} K \chi^2 + S(\chi) \beta^2\,,
    \label{eq:nuclear_matter_exp}
\end{align}
where $\epsilon$ is the baryonic energy density and $\chi = \frac{(n-n_0)}{3 n_0}$ is the relative density difference to saturation density.
The expansion coefficients (called nuclear matter properties) for symmetric nuclear matter with $\beta=0$ are the binding energy $B$ and the incompressibility $K$.
The third term in \cref{eq:nuclear_matter_exp} depends on the isospin asymmetry and is called symmetry energy.
It is defined via the second derivative of the energy per particle with respect to $\beta$ and can be expanded as
\begin{equation}
    S(\chi) = \left. \frac{1}{2} \del{^2E/A}{\beta^2} \right|_{\beta=0} \approx
        E_{\rm sym} + L \chi\,.
        \label{eq:Esym-expand}
\end{equation}
with the symmetry energy at saturation density $E_{\rm sym}$ and the slope parameter $L$, which define two nuclear matter properties for neutron matter ($\beta=1$).
In this quadratic approximation, the symmetry energy is the difference between the energy per particle of neutron matter and symmetric nuclear matter:
\begin{align}\label{eq:Esym}
 S(\chi) = \frac{E}{A}(\beta=1) - \frac{E}{A}(\beta=0)\,.
\end{align}
It has been shown in microscopic calculations that the quadratic expansion of the energy per particle provides a good description of the EOS around saturation density \ceg{Drischler2016a, Wellenhofer2016a, Somasundaram2021a, Keller2023a}, with non-quadratic contributions arising mainly from the kinetic energy density that is fully included in EOS functionals.

As in \citet{Yasin2020a}, we use the open-source SROEOS code \citep{Schneider2017a, Schneider2018a}, which is based on a Skyrme density functional plus liquid drop model to describe the low-density EOS.
The nuclear matter parameters $B$, $K$, $E_{\rm sym}$, and $n_0$ are systematically varied by adjusting the parameters of the Skyrme density functional while the effective mass $m^*$ at saturation density is varied directly.
To explore the impact of these parameters we use nine \ac{EOS} tables.
Eight of the tables were created with the SROEOS code, six of which were developed and used in \citet{Yasin2020a} while two are used for the first time in this work.
Finally, the ninth \ac{EOS} is the Shen \ac{EOS} \citep{Shen1998a}.
The nuclear matter properties of these \ac{EOS} models are summarized in \cref{tab:eos_table}.

\begin{table*}
  \centering
  \caption{%
    Nuclear matter properties for all employed \acp{EOS} models, following the notation in \citet{Yasin2020a} Note that $m^*$ is the effective nuclear mass at saturation density.
    Also given are the pressure and the dimensionless slope of the pressure at saturation density for $T=0$ and $Y_e=0.1$.
    \label{tab:eos_table}}
  \begin{tabular}{lcccccccc}
    \toprule
    EOS     & $m^*$ & $B$  & $K$  & $E_{\rm sym}$  & $L$  & $n_0$ & $P|_{n=n_0,T=0,Y_e=0.1}$ & $\left. \frac{n}{P} \del{}{n} P \right|_{n=n_0,T=0,Y_e=0.1}$ \\
    & [$m_n$] & [MeV] & [MeV] & [MeV] & [MeV] & [fm$^{-3}$] & [MeV fm$^{-3}$] & \\

    \midrule
    \LSl   & 1.0   & 16.0 & 175 & 29.3 &  73.7 & 0.155 & 3.17 & 2.77 \\
    \LSm   & 1.0   & 16.0 & 220 & 29.3 &  73.7 & 0.155 & 3.19 & 3.03 \\
    \LSh   & 1.0   & 16.0 & 255 & 29.3 &  73.7 & 0.155 & 3.19 & 3.23 \\
    \msm   & 0.8   & 16.0 & 220 & 29.3 &  79.3 & 0.155 & 3.39 & 3.07 \\
    \msS   & 0.634 & 16.0 & 220 & 29.3 &  86.5 & 0.155 & 3.66 & 3.13 \\
    \msKS  & 0.634 & 16.0 & 281 & 29.3 &  86.5 & 0.155 & 3.68 & 3.43 \\
    \msKES & 0.634 & 16.0 & 281 & 36.9 & 109.3 & 0.155 & 4.45 & 3.18 \\
    SkShen & 0.634 & 16.3 & 281 & 36.9 & 109.4 & 0.145 & 5.11 & 3.18 \\
    Shen   & 0.634 & 16.3 & 281 & 36.9 & 110.8 & 0.145 & 4.98 & 3.14 \\
    \bottomrule
  \end{tabular}
\end{table*}

With each of the \ac{EOS} models, we systematically vary one nuclear matter property at a time.
Following \citet{Yasin2020a}, we use the model \LSm as the fiducial model which is based on the same Skyrme parametrization as the LS220 \ac{EOS}.
Its incompressibility parameter is $K = \unit[220]{MeV}$ and its effective mass is not density-dependent and simply given by the neutron mass $m^* = m_n = \unit[939.5654]{MeV}$.
First, we vary the $K$ parameter to 175 and \unit[255]{MeV}, resulting in the \ac{EOS} \LSl and \LSh, respectively.
These values represent the upper and lower bounds predicted by chiral effective field theory calculations \citep{Hebeler2011a, Drischler2016a, Drischler2019a}.
Second, we change the nucleon effective mass at saturation density to $m^* = 0.8 m_n$ (\msm) and $m^* = 0.634 m_n$ (\msS).
The value of $m^* = 0.634 m_n$ is chosen because it is used in the Shen \ac{EOS}.
Finally, starting from \msS, we also change the remaining nuclear matter properties to match the values of the Shen \ac{EOS}, starting with $K=\unit[281]{MeV}$, followed by the symmetry energy $E_{\text{sym}}=\unit[36.9]{MeV}$, and the value of the saturation density $n_0= \unit[0.145]{fm^{-3}}$ and the binding energy $B=\unit[16.3]{MeV}$.
This results in the \acp{EOS} \msKS, \msKES, and SkShen, respectively.
We include the original Shen \ac{EOS} for comparison.

\begin{figure}
  \centering
  \includegraphics{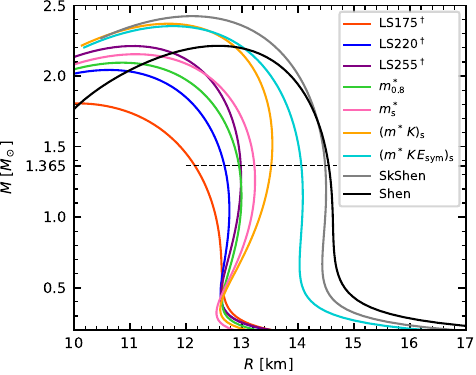}
  \caption[The mass-radius relation of cold non-rotating \acsp{NS} for the \acsp{EOS} used in this paper]{%
    The mass-radius relation of cold non-rotating \acp{NS} for the \acp{EOS} used in this paper.
    The horizontal dashed line marks $1.365 M_\odot$ which is the mass of the \acp{NS} explored in the simulations in this work.
    \label{fig:MR}}
\end{figure}

\Cref{fig:MR} shows the mass-radius relation of cold  \acp{NS} for all of the \acp{EOS}.
Changing the incompressibility has a large influence on the radius of \acp{NS} with high masses (compare \LSl, \LSm, and \LSh or \msS and \msKS in \cref{fig:MR}).
The effective mass impacts intermediate to high mass \acp{NS} but has less of an effect at high masses compared to the incompressibility.
Varying the symmetry energy mostly has a large effect on the radii of low mass \acp{NS}.
The horizontal dashed line marks the \acp{NS} with a mass of $1.365 M_\odot$.
These are the radii of the \acp{NS} in the initial data of the simulations (see \cref{sec:overview}).

\begin{figure}
  \centering
  \includegraphics{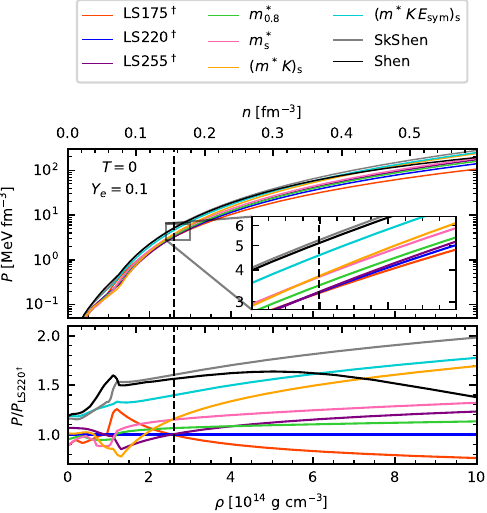}
  \caption{%
    Pressure versus density at $T=0$ and $\ye = 0.1$ for the \acp{EOS} used in this work.
    The lower panel shows the curves normalized to the pressure in the fiducial model \LSm for comparison. The vertical dashed lines mark the saturation density 0.155\,fm$^{-3}$.
    \label{fig:pressure}}
\end{figure}

\Cref{fig:pressure} shows the pressure versus density at $T = 0$ and $\ye = 0.1$ for all the employed \acp{EOS}.
The pressure is related to the derivative of the energy density with respect to the particle density. Since the pressure of symmetric matter vanishes at saturation density, the slope of the pressure at $\beta=0$ is governed by the incompressibility.
Increasing $K$ leads to larger pressure at densities above $n_0$, but also smaller pressures below $n_0$ (compare \LSl, \LSm, and \LSh in \cref{fig:pressure}).
Changing the symmetry energy also changes the slope parameter $L$, which characterizes the pressure of neutron matter.
Thus, it has a relatively large effect close to saturation density but smaller at higher densities (compare \msKS and \msKES in \cref{fig:pressure}).
Finally, the effective mass influences the degeneracy pressure from the kinetic energy density, which for our Skyrme density functional is given by \citep{Lattimer1991a}
\begin{align}
  \label{eq:deg_en}
  P_{\text{deg.}} (T=0) = \sum_t \left(\frac{5}{3} \frac{1}{2 m^*} - \frac{1}{2 m} \right)
   \frac{(3 \pi n_t)^{5/3}}{5 \pi^2}\,.
\end{align}
Since the effective mass enters the degeneracy pressure inversely, decreasing the effective mass increases the cold pressure.
Furthermore, for the \acp{EOS} used here reducing the effective mass implicitly increases the $L$ parameter due to the way that the Skyrme parameters are determined.
Therefore, the pressure is increased for the full range of densities in the \ac{NS} ($\rho > \dens{14}$).

While increasing $K$ and reducing $m^*$ both generally increase the pressure (\ie, the stiffness of the \ac{EOS}), the two parameters have different effects around saturation density.
This can be visualized by considering the pressure and the dimensionless slope of the pressure $\left. \frac{n}{P} \, \del{}{n}P \right|_{n=n_0,T=0,Y_e=0.1}$ at saturation density for $T=0$ and $Y_e=0.1$.
\begin{figure}
  \centering
  \includegraphics{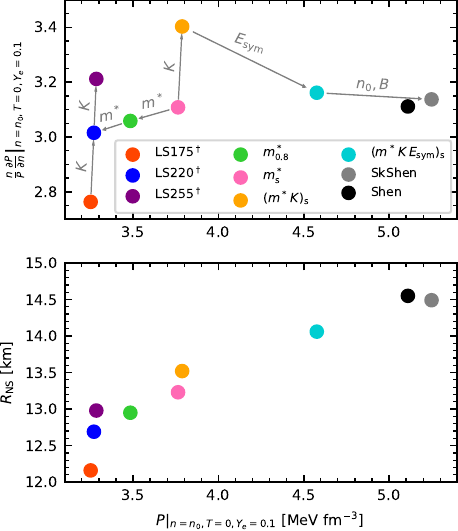}
  \caption{%
    Top panel: the pressure and the dimensionless slope of the pressure at saturation density for $T=0$ and $Y_e=0.1$.
    Arrows show the relevant nuclear matter property changes.
    Lower panel: the pressure at saturation density for $T=0$ and $Y_e=0.1$ and the radius of the initial \acp{NS}.
    \label{fig:P_dPdn}}
\end{figure}
Both quantities are shown in the top panel of \cref{fig:P_dPdn} and given in \cref{tab:eos_table} for all \ac{EOS} models considered in this work.
As discussed above, $K$ only affects the slope of the pressure at saturation density but not the value of the pressure itself.
Reducing the effective mass (\LSm, \msm, \msS), increases the pressure at saturation density and its slope but to a lesser degree than increasing $K$.
This is due to an increase in the next-order term in the symmetry energy $K_{\text{sym}}$.
Finally, \msKES, SkShen, and Shen exhibit a higher pressure at saturation density, but their dimensionless pressure slope is comparable to that of \msS.
The radii of the initial \acp{NS} are shown in the lower panel of \cref{fig:P_dPdn}.
They are mostly correlated with the pressure at saturation density.

The total energy density and pressure are often divided into a cold and thermal part (indicated by the subscripts $\text{cold}$ and $\text{th}$, respectively).
\begin{equation}
  \label{eq:cold_th}
  P(n, T) = P_{\text{cold}}(n, T=0) + P_{\textnormal{th}}(n, T)\,,
\end{equation}
where $P_{\textnormal{th}}(n, T=0) = 0$.
The temperature dependence is often described by the thermal index $\Gamma_{\textnormal{th}}$ given by \ceg{Bauswein2010a}
\begin{equation}
  \label{eq:gammath}
  \Gamma_{\textnormal{th}} = 1 + \frac{P_{\textnormal{th}}}{\epsilon_{\textnormal{th}}}
  = 1 + \frac{P - P_{\text{cold}}}{\epsilon - \epsilon_{\text{cold}}}\,.
\end{equation}
\begin{figure}
  \centering
  \includegraphics{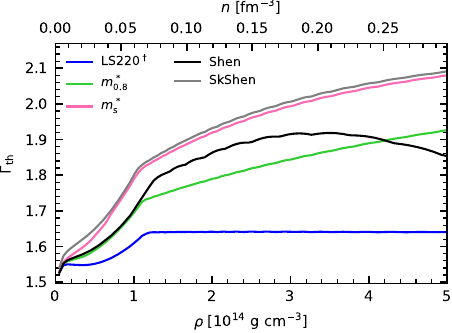}
  \caption[Thermal index $\Gamma_{\textnormal{th}}$ for \LSm, \msm, and \msS]{%
    Thermal index $\Gamma_{\textnormal{th}}$ for the \LSm, \msm, \msS, SkShen, and Shen \acp{EOS}.
    SkShen exhibits a slightly different $\Gamma_{\textnormal{th}}$ compared to \msS due to the different saturation density.
    Although the saturation density and the effective mass at saturation density are the same in SkShen and Shen, their thermal index is different due to the underlying relativistic mean field functional in Shen.
    \label{fig:gammath}}
\end{figure}%
For a non-interacting gas of non-relativistic fermions with density-dependent effective mass, the thermal index is determined by the density dependence of the effective mass \ceg{Constantinou2015a, Lattimer2016a, Keller2021a}:
\begin{equation}
    \Gamma_{\textnormal{th}} = \frac{5}{3} - \frac{n}{m^*} \frac{\partial m^*}{\partial n}\,,
    \label{eq:gammath_approx}
\end{equation}
Chiral effective field theory calculations have found that Eq.~\eqref{eq:gammath_approx} is a good approximation for the full thermal index via Eq.~\eqref{eq:gammath} \citep{Carbone2019a, Keller2021a, Keller2023a}.
In \cref{fig:gammath}, the thermal index inside the \acp{NS} is shown versus density for the three values of the effective mass used in our simulations, and for the Shen \ac{EOS}.
For the non-relativistic Skyrme density functional based EOSs, the thermal index is the same using Eq.~\eqref{eq:gammath} or Eq.~\eqref{eq:gammath_approx}.
Therefore, lowering the effective mass at saturation density increases the thermal index.
We also note that this behavior is due to the parametrization of the density-dependence of the effective mass, which decreases with increasing density, leading to an increasing thermal index.
However, the results from \citet{Carbone2019a, Keller2021a, Keller2023a} show that $\Gamma_{\textnormal{th}} (n)$ has a maximum and starts to decrease at higher density, which originates from a minimum of the effective mass around saturation density (see also \citet{Huth2021a}).
Note that the thermal index of the Shen EOS does not follow the expression from Eq.~\eqref{eq:gammath_approx} due to relativistic mean field functional used.

Several previous works investigated thermal effects in binary compact object mergers in isolation by changing the thermal pressure independently from the cold \ac{EOS} \ceg{Bauswein2010a, Hotokezaka2013a, Raithel2021a, Fields2023a}.
The setup of \citet{Fields2023a} is also based on variations of the effective mass with the SROEOS code and is very similar to ours.
However, by employing a functional with more free parameters than ours, they also constrain the $L$ parameter and the pressure at four times $n_0$.
Therefore, the variations of the effective mass in their work only change the thermal pressure and do not affect the \ac{EOS} at zero temperature, while in our EOS function both the cold and the thermal pressure are changed by varying the effective mass.


\section{Merger and post-merger dynamics} \label{sec:remnant}

Many of the observable features of a \ac{BNS} merger depend sensitively on the dynamics during the first \unit[10$-$20]{ms} after the merger.
Therefore, we investigate this early post-merger phase in this section.
We specifically focus on the effect of the \ac{EOS} on the collapsing behavior, the initial contraction and deformation of the massive \acp{NS}, the role of thermal effects, and the formation of the disc.

\subsection{Evolution of the central object}\label{sec:contraction}
\begin{figure}
  \centering
  \includegraphics{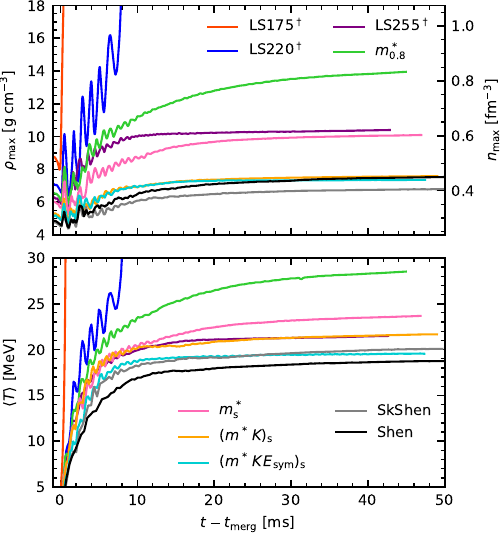}
  \caption{%
    Maximum density (upper panel) and average remnant temperature (lower panel) versus time.
    \label{fig:NS_stuff}}
\end{figure}

\Cref{fig:NS_stuff} shows the time evolution of the maximum density inside the remnant \ac{NS}.
As the two cores of the original \acp{NS} combine after the merger, the maximum density rises.
Consequently, the pressure in the \ac{NS} core increases as well.
If the pressure given by the \ac{EOS} is relatively low (\ie, for a soft \ac{EOS}) a prompt collapse occurs, as is the case of model \LSl.
The other models initially form a massive \ac{NS} remnant.
Note that the pressure at high densities (several times the saturation density) is relevant for the collapse.
In the context of the quantities introduced in \cref{sec:eos_tables} and shown in \cref{fig:P_dPdn}, the pressure at high density is determined by a combination of the pressure and its slope at saturation density.

All models except \LSl do not promptly collapse.
In these cases, the cores of the initial \acp{NS} bounce off each other initially.
Thus, in the center of the newly formed massive \ac{NS}, two density maximums are present and form an oscillating bar shape.
The oscillations can be seen in the maximum density in \cref{fig:NS_stuff}.
It typically takes \unit[10$-$20]{ms} for the \ac{NS} cores to merge.
During this process, the central density and consequently the pressure in the \ac{NS} core keeps increasing.
The contraction ends either due to the delayed formation of a \ac{BH} or when the pressure becomes large enough to balance the central object's self-gravity.
In this case, a meta-stable \ac{NS} remains in the center.
For \LSm, several oscillations with increasing amplitude occur before the central object collapses to a \ac{BH}, at roughly \unit[8]{ms} after the merger.
Because these models exhibit a larger slope of the pressure as a function of density, the pressure rises faster as the central density increases.
In \cref{fig:NS_stuff}, one can see that the central density of models with larger incompressibilities (\LSh, \msKS, \msKES) stops rising already after \unit[10$-$15]{ms} while it keeps increasing for more than \unit[20]{ms} in \msm, \msS.
However, this trend does not hold for the models SkShen and Shen in which the contraction of the remnant continues for longer times as well.
The central density at the end of the non-collapsing simulations seems to be determined by the pressure at $6-7 \times \dens{14}$ (compare \cref{fig:pressure} with \cref{fig:NS_stuff}).

\begin{figure}
  \centering
  \includegraphics{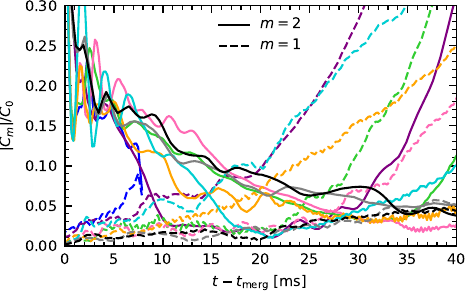}
  \caption{%
    Complex azimuthal $m = 1$ and $2$ decomposition of the density in the $xy$-plane calculated with \cref{eq:decomp}.
    Solid and dashed lines correspond to the $m=2$ and $m=1$ modes, respectively.
    The colors denote the different models and have the same meaning as in \cref{fig:NS_stuff}.
    The transition to a dominant $m=1$ mode happens first for \LSl, followed by \msKES, and \msKS.
    \label{fig:xy_decomp}}
\end{figure}

During this initial phase of contraction, the remnant is highly deformed and oscillates, where the dominant deformation mode is a $m=2$ bar-shaped deformation.
However, after some time, the $m=1$ deformation becomes dominant.
This change influences the gravitational wave emission, disk formation, and matter ejection \citep{Bernuzzi2014a, Kastaun2015a, Paschalidis2015a, East2016a, East2016b, Lehner2016a, Radice2016a, Nedora2019a, Nedora2021b}.
In \cref{fig:xy_decomp}, the $m=1$ and $2$ deformation calculated with \cref{eq:decomp} is shown.
The models undergo the change from a dominating $m=2$ to a dominant $m=1$ mode at different times.
In \LSh, \msKS, and \msKES, the transition occurs already after roughly 9, 15, and \unit[17]{ms}, respectively, which is significantly earlier compared to the other models.
This roughly matches the time at which the contraction of the massive \acp{NS} in these models stops, indicating that the stability of the initial bar-shaped mode might be linked to the contraction of the remnant and therefore to the slope of the pressure.

\subsection{Thermal effects}\label{sec:thermal-effects}
\begin{figure*}
  \centering
  \includegraphics{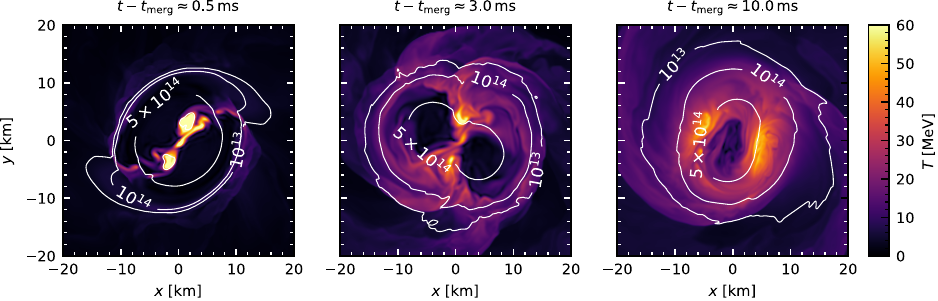}
  \caption{%
    Temperature profiles inside the remnant for simulation \msm at approximately $0.5$, $3$, and \unit[10]{ms} post-merger.
    Density contours of $10^{13}$, $10^{14}$, and $5 \times \dens{14}$ are show in white.
    \label{fig:temp_profile}}
\end{figure*}

\begin{figure}
  \centering
  \includegraphics{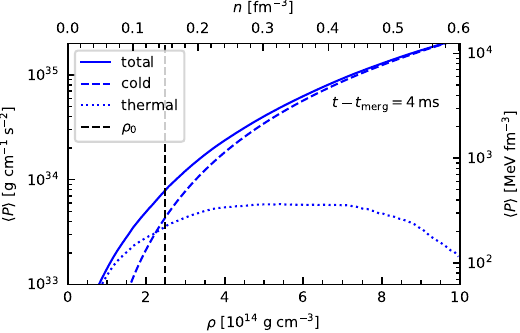}
  \caption{%
    Average thermal, cold, and total pressure over the matter density for simulation \LSm, roughly \unit[4]{ms} after the merger.
    The thermal contributions to the pressure become relevant for densities close to the saturation density.
    \label{fig:press_th_comp}}
\end{figure}

During the merger, the initially cold \ac{NS} matter is heated to several tens of $\textnormal{MeV}$ or more by shocks originating from the interface of the collision.
The high-density \ac{NS} cores remain comparatively cool because the shocks do not penetrate them.
As the colder cores merge, the hot matter is redistributed into a ring shape at densities of approximately $1 - 5 \times \dens{14}$.
This evolution can be seen in \cref{fig:temp_profile}, which shows temperature profiles inside the remnant for simulation \msm at approximately $0.5$, $3$, and \unit[10]{ms} post-merger.
Because the center of the \ac{NS} is cold and very dense, the cold pressure dominates and thermal contributions to the pressure are not relevant in this region.
However, the thermal pressure plays a significant role in the hot ring close to saturation density and even becomes larger than the cold pressure in the outer layers of the \ac{NS} as can be seen in \cref{fig:press_th_comp}.
Thermal effects are therefore especially important for disk formation and matter ejection but their impact on the post-merger gravitational wave emission is small \ceg{Bauswein2010a, Hotokezaka2013b}.

Generally, a stiffer \ac{EOS} leads to less shock-heating due to two effects.
Larger initial \ac{NS} radii lead to an earlier merger because the \acp{NS} come into contact at larger separations.
This leads to smaller orbital and radial velocities \ceg{Radice2020a} and thus a less intense initial bounce and less shock-heating.
Furthermore, with a softer \ac{EOS}, the remnant will contract more before the first bounce occurs, which leads to stronger oscillations in the early post-merger phase.

Apart from the temperature, the thermal pressure depends on the thermal index which is determined by the density dependence of the effective mass in our EOS models.
Reducing the effective mass at saturation density increases the thermal index and therefore the effectiveness of shock heating \citep{Hotokezaka2013a}.
However, reducing the effective mass in our \ac{EOS} setup also increases the $L$ and $K_{\rm sym}$ parameters and thus the cold pressure, which in turn leads to a less violent merger.
Therefore, the effect of changing the thermal index cannot be observed independently in our simulations.

The lower panel of \cref{fig:NS_stuff} shows the evolution of the average temperature inside the remnant as a function of time.
In \LSm the remnant is heated up much more than in the non-collapsing cases due to the increasingly violent oscillations.
Because stiffer \acp{EOS} result in a less violent merger and reduced shock heating, non-collapsing remnants with larger maximum densities generally also have larger temperatures.
However, there are some deviations from this general trend.

First, the average remnant temperature depends mostly on the outer layers of the \ac{NS} (and thus on the \ac{EOS} close to saturation density) since this is the region that is heated by shocks.
Furthermore, the initial \ac{NS} radii (and therefore the radial infall velocities of the \acp{NS} at merger) depend on the \ac{EOS} at lower densities as well.
The maximum density, however, is more sensitive to the pressure at high densities.
This can be seen by comparing \msS and \msKS, which have the same pressure at saturation but differ at high density.
Consequently, the maximum density is lower in \msKS but the average temperatures are very similar.
Similarly, \msKS and \msKES show a very similar evolution of the central density but the average temperature of \msKES is lower because the two \acp{EOS} differ only at low densities.

Second, the thermal index influences the effectiveness of shock heating which can be seen by comparing \LSh and \msS.
The \LSh \ac{EOS} exhibits lower pressures for all densities compared to \msS (see \cref{fig:pressure}).
Therefore, the central density in \LSh is higher than in \msS.
Nonetheless, its average remnant temperature is lower because \msS has a higher thermal index which increases the efficiency of the shock heating.

\subsection{Comparison of Shen and SkShen}
The SkShen \ac{EOS} matches the effective mass, incompressibility, symmetry energy, binding energy, and saturation density of the Shen \ac{EOS}.
The two \ac{EOS} models are thus similar close to saturation density but differ more and more at higher densities.
Furthermore, the thermal index of the Shen \ac{EOS} is lower compared to SkShen.
This is expected because matching the effective mass at saturation density does not reproduce the thermal index if the density dependence of the effective mass is different.

Due to the increased pressure at high densities, the central density in the model Shen is larger than in SkShen.
The average remnant temperature, however, is larger in SkShen because it has a larger thermal index, resulting in more effective shock heating.
Nonetheless, considering the large deviation of the two \acp{EOS} at high densities, the evolution of the remnants in the models SkShen and Shen are remarkably similar.
This indicates, that the post-merger evolution is largely determined by the \ac{EOS} around saturation density, even though the maximum density in SkShen reaches up to 3 times the saturation density.

\subsection{Accretion disk evolution}
\label{sec:disk}

\begin{figure}
  \centering
  \includegraphics{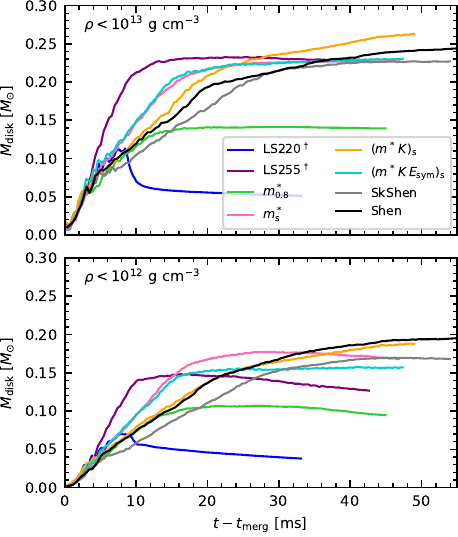}
  \caption{%
    Evolution of the disk mass for all simulations except \LSl.
    The top and bottom panels correspond to different definitions of the disk boundary.
    \label{fig:disk_mass}}
\end{figure}

We define the disk mass as the total baryon rest mass outside of the \ac{NS}, \ie, the region with $\rho < \rho_{\text{disk}} = \dens{13}$.
However, this definition for the \ac{NS} surface is relatively arbitrary.
Specifically, it includes the transition region between the disk and \ac{NS}.
Thus, we also calculate all disk-related quantities for $\rho_{\text{disk}} = \dens{12}$ which excludes this transition region.
Therefore, the second definition offers a more conservative estimate of which matter contributes to the disk.
\Cref{fig:disk_mass} shows the evolution of the disk mass for all simulations except \LSl.
The left and right panels correspond to $\rho_{\text{disk}} =  10^{12}$ and $\rho_{\text{disk}} = \dens{13}$, respectively.
The disk masses at the end of the simulations are listed in \cref{tab:overview}.

The formation of the accretion disk depends on the \ac{EOS} properties due to multiple effects.
The most important factor for the disk mass is the fate of the merger remnant.
In simulation \LSl, the central object collapses to \ac{BH} almost immediately after the merger, so no disk is formed, while the \ac{BH} formation in \LSm is delayed long enough for a disk to form.
However, roughly half of the disk mass is swallowed upon collapse.
\begin{figure*}
  \centering
  \includegraphics{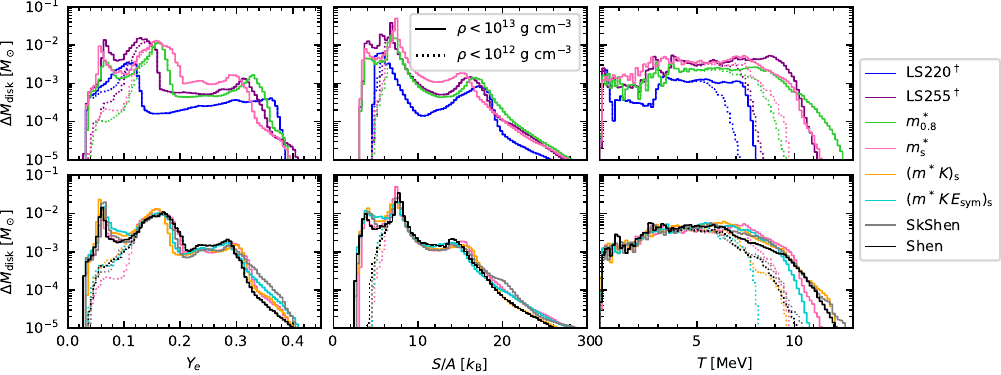}
  \caption{%
    Mass-weighted histogram of the electron fraction, entropy per baryon, and temperature in the disk at $t = \unit[30]{ms}$.
    Dotted lines show the histograms when matter with $\rho > \dens{12}$ is excluded.
    The EOS models are split into two panels for better readability.
    \label{fig:disk_hist}}
\end{figure*}
For non-collapsing mergers, the relation of the disk mass to the \ac{EOS} is more complicated.
On one hand, the disk mass originating from the tidal disruption of the \acp{NS} is larger for stiffer \acp{EOS}.
For a softer \ac{EOS}, on the other hand, matter ejection due to shock heating is enhanced.
Furthermore, the remnant is more compact and rotates faster \ceg{Bernuzzi2020a, Nedora2021b}, which also increases the disk mass.
In addition to the \ac{EOS} at zero temperature, thermal effects have to be considered.
A larger thermal index enhances the amount of matter ejected to the disk because of shock heating \citep{Hotokezaka2013a}.
It is thus hard to find a correlation between the disk mass and specific nuclear matter properties.

At the end of the simulations that do not form a \ac{BH} the disk mass is close to $0.25 M_\odot$.
The only exception is \msm with a disk mass of $0.14 M_\odot$.
The mass of the disk in \LSh rises very fast in the first \unit[10]{ms} after the merger.
Furthermore, the disk is particularly massive in comparison to the other models, considering that \LSh is comparable to \msm in terms of stiffness.
This changes if the inner boundary of the disk is defined by $\rho < \dens{12}$.

Even though the evolution of the disk mass and angular momentum varies for the different \acp{EOS}, the composition and structure of the disk are relatively similar in all non-collapsing simulations.
The mass-weighted histogram of the electron fraction, entropy per baryon, and temperature in the disk at $t = \unit[30]{ms}$ is shown in \cref{fig:disk_hist}.
All histograms show a triple peak structure, typical for equal mass \ac{BNS} mergers \citep{Nedora2021b}.
The peak at the lowest entropy and electron fraction corresponds to the interface between the remnant and the disk.
The dotted lines in \cref{fig:disk_mass} show the histograms for the matter at densities below $\dens{12}$ which lack this high temperature, low entropy, and low electron fraction tail.

The second peak at $S/A \approx 4-8\,k_{\text{B}}$ and $\ye \approx 0.1-0.2$ represents the bulk of the disk.
The disk bulk has densities between $10^{10} - \dens{12}$ and extends to roughly \unit[100]{km}.
For most simulations, the main peak in the electron fraction and entropy are located at $\ye = 0.16$ and $S/A = 7.5\,k_{\text{B}}$.
The largest deviations from this trend are \LSm and \LSh.
Since the central object in \LSm is a \ac{BH}, the hot and dense interface to the \ac{NS} is missing, so there are no spiral arms in the disk.
Furthermore, the strong neutrino irradiation from the central \ac{NS} is also missing.
Therefore, the disk has lower temperatures, entropies, and electron fractions.
Due to the higher incompressibility of \LSh, a larger fraction of the disk is located closer to the massive \ac{NS}.
This effectively shifts the average bulk entropy and energy toward lower values.
The same is true for \msKS and \msKES, however, to a lesser extent.

The third peak with high electron fractions and entropies is due to the matter outside of the main bulk of the disk.
During the disk evolution, this matter becomes partially unbound.
Thus, its composition matches that of the disk wind ejecta in \cref{fig:ejecta_hist} (see \cref{sec:ejecta_properties} for an extended discussion).


\section{Ejecta properties}
\label{sec:ejecta_properties}
\begin{figure}
  \centering
  \includegraphics{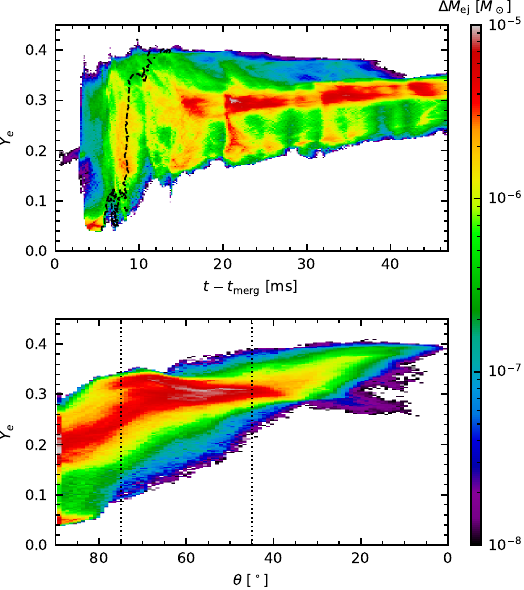}
  \caption{%
    Mass-weighted histograms of ejecta properties in model \msS.
    Top panel: Two-dimensional histogram of the electron fraction versus the time after the merger at which the ejecta cross the detection sphere at \unit[300]{km}.
    The dashed contour marks the transition point from dynamical ejecta to shock-heated ejecta and the horizontal dashed line shows the cut at $\ye = 0.1$ used to distinguish the tidal ejecta.
    Bottom panel: Two-dimensional histogram of the electron fraction versus the polar angle.
    \label{fig:ejecta_2D_hist}}
\end{figure}
\begin{figure*}
  \centering
  \includegraphics{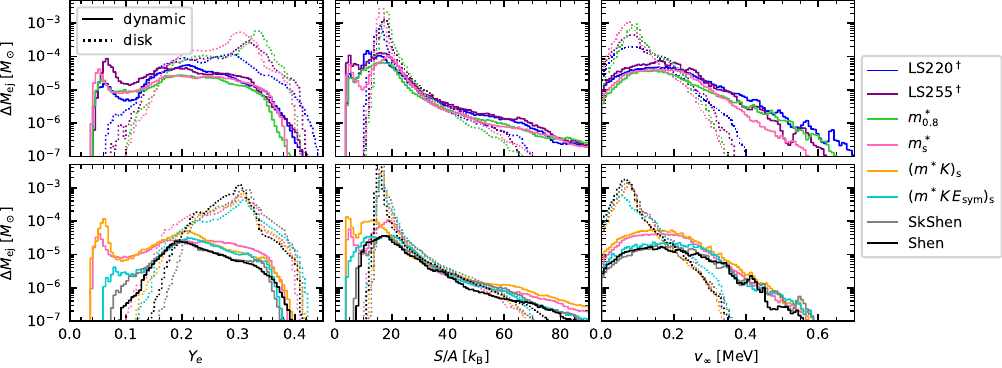}
  \caption{%
    Mass-weighted histogram of the electron fraction, entropy per baryon, and asymptotic velocity of the ejected matter.
    Solid and dotted lines represent dynamical ejecta and disk ejecta, respectively.
    The EOS models are split into two panels for better readability.
    \label{fig:ejecta_hist}}
\end{figure*}

The mass ejection in \ac{BNS} mergers occurs through multiple channels.
These can be separated into two categories: dynamic and disk (or secular) ejecta.
The former are expelled within a few milliseconds after the merger, while the latter are ejected over timescales of \unit[10]{ms} up to \unit[10]{s} \ceg{Fujibayashi2023a}.

As described in \cref{sec:postprocessing}, we define the dynamical ejecta as matter that fulfills the geodesic criterion, while matter that fulfills the Bernoulli criterion but not the geodesic criterion is associated with the disk ejecta.
The same division is used in \citet{Nedora2021b, Combi2023a}.
Note, however, that this division scheme is only approximate as the transition from dynamical ejecta to disk ejecta happens continuously between \unit[5-10]{ms} after the merger.
Furthermore, we separate the dynamical ejecta into the tidal and shock-heated components.
For this, we simply count dynamically ejected fluid elements toward the tidal component if their electron fraction is smaller than 0.1 and toward the shock-heated component otherwise.
The classification into dynamical and disk ejecta and tidal and shock-heated ejecta is made on the detection sphere at \unit[300]{km} radius.
  Note that this distinction based on the electron fraction only works, if the ejecta extraction radius is small enough.
  At later times, neutrino captures increase the electron fraction of the tidal ejecta making it more difficult to distinguish them from the shock-heated ejecta.

In \cref{fig:ejecta_2D_hist}, we examine the detailed ejecta composition for the simulation \msS.
We use \msS as reference here but all other simulations show similar trends to the ones discussed below.
The left panel shows the 2D histogram of the electron fraction of the ejected fluid elements versus the time at which they reach \unit[300]{km}.
The right panel shows the histogram of their electron fraction versus the polar angle of ejection $\theta$.
The division into dynamical and disk ejecta components is approximately shown by the dashed contour in the left panel.
The dashed horizontal line shows the cut at $\ye = 0.1$.
Tidally ejected fluid elements are clearly visible in the lower left corner of both panels (\ie, at low $\ye$, early times, and in the equatorial direction).
However, the distinction between the shock-heated and disk ejecta is less clear.
We observe roughly separate bursts of mass ejection at $t - t_{\text{merg}} \approx \unit[8, 14, 20, \dotsc]{ms}$ in the left panel.
The first burst is the shock-heated ejecta, while the later bursts are caused by the oscillations following the initial bounce of the newly formed massive \ac{NS}.
Especially in the case of \msS, the division based on the geodesic criterion does not properly separate the first burst from the subsequent ones and slightly underestimates the shock-heated ejecta mass.

\Cref{fig:ejecta_hist} shows the accumulated mass histogram of the ejecta electron fraction, entropy per baryon, and asymptotic velocity.
Solid lines show the dynamical ejecta and dotted lines the disk ejecta.
If the tidal ejecta do not experience significant shock heating, they consist of almost pure cold \ac{NS} matter ($\ye < 0.1$) and have low entropies ($S/A < 10 k_{\text{B}}$).
This part forms a peak at low electron fractions and entropies in \cref{fig:ejecta_hist}.
However, a fraction of the tidally removed matter can be reprocessed by shocks and thus counts towards the shock-heated ejecta component in our classification scheme.
In \msKES, SkShen, and Shen, this is the main part of the tidal ejecta.

The shock-heated ejecta reach much higher temperatures and entropies compared to the tidal ejecta.
Therefore, positron and electron-neutrino captures increase the electron fraction of the matter to $\ye \approx 0.1 - 0.4$.
A small part of the shock-heated ejecta is responsible for the high-velocity and high-entropy tail in the right panels of \cref{fig:ejecta_hist}.
These ejecta might lead to additional observable features in the kilonova and is larger for softer \acp{EOS} \cite[see, \eg,][and references therein]{Dean2021a}.

\begin{figure}
  \centering
  \includegraphics{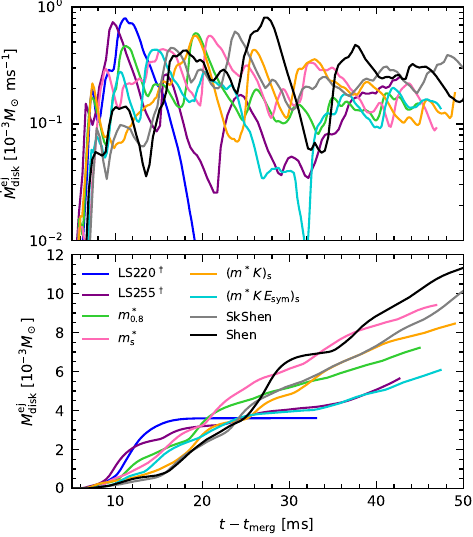}
  \caption{%
    Mass ejection rate and total ejected mass of the disk ejecta component as a function of time.
    \label{fig:ejecta_mass}}
\end{figure}

The evolution of the mass ejection rate and the integrated total ejected mass for the disk component is shown in \cref{fig:ejecta_mass}.
On the timescales of our simulations, the mass ejection in the early post-merger phase is primarily driven by the oscillating double-core structure in the massive \ac{NS}.
With each bounce, matter becomes unbound as the central density reaches a minimum \citep{Nedora2019a, Nedora2021b, Combi2023a}.
The left panel of \cref{fig:ejecta_2D_hist} shows several such bursts of matter ejection after $t \approx \unit[10]{ms}$.
Furthermore, energy deposition by neutrino absorption and scattering \citep{Dessart2009a, Perego2014a, Just2015a, Radice2018a} enhances the ejection matter from the early post-merger remnant.
In \LSm, the hypermassive \ac{NS} collapses and thus no longer provides an efficient mechanism for the ejection of matter through neutrino irradiation and the oscillation of the \ac{NS}.
Thus the ejection of matter stops after $t > \unit[20]{ms}$.
For all non-collapsing models, however, the ejection of matter is still ongoing at the end of the simulation and viscous effects will eject a 10\%$-$50\% of the disk mass on timescales of seconds \ceg{Fernandez2013a, Just2015a, Fujibayashi2018a, Fujibayashi2020a, Fahlman2022a, Hayashi2022a}.
Thus, the total disk ejecta mass is not converged in our simulations.

We find that the mass ejection in the disk phase is more effective while the $m = 2$ bar-shaped deformation of the remnant drives two spiral arms into the disk (see \cref{fig:xy_decomp}).
The model \LSh exhibits a period of low mass ejection after $\sim \unit[15 - 20]{ms}$ that lasts approximately \unit[20]{ms}.
A similar feature is visible for \msKES after $\sim$ \unit[25]{ms}, lasting approximately \unit[10]{ms}.
Both correlate well with periods of low $m = 2$ deformations of the remnant with a delay of $\sim$ \unit[10]{ms} which is roughly the time it takes ejecta to travel to the detection radius at \unit[300]{km}.
Later, a one-armed spiral wave, driven by the $m=1$ deformation, appears, resulting in an increased mass ejection rate.
Similar results were found by \citet{Nedora2019a, Nedora2021b}.

Directly above the massive \ac{NS} the neutrino heating is very strong.
However, the baryonic matter density in the polar direction is low, so the neutrino heating unbinds only a small amount of matter in this region.
This matter is ejected toward the polar direction with relatively high velocities and electron fractions.

The neutrino heating in the diagonal direction enhances the mass outflow significantly and increases its electron fraction.
This is visible in \cref{fig:ejecta_2D_hist}:
in the right panel, the angles $\theta \approx 45 - 75^\circ$ are marked by dotted lines and encompass the largest ejecta component (the roughly horizontal bar-shaped region).
In the left panel, this component can be seen after $\unit[15-20]{ms}$ at $\ye \approx 0.3$.
This part of the ejecta forms a peak at $\ye \approx 0.3$ in \cref{fig:ejecta_hist} for all models except \LSm, where it is missing due to the lack of neutrino emission from the collapsed massive \ac{NS}.
\Cref{fig:ejecta_2D_hist} also shows how this component evolves.
With time the disk becomes thinner and more transparent to neutrinos.
As a result, the neutrino-irradiation intensifies and the outflow becomes more proton-rich and moves closer to the equatorial plane.

Initially, the oscillations of the massive \ac{NS} ejects matter also in the equatorial direction.
The dense bulk of the disk shields the matter in the equatorial plane outside the disk from neutrino irradiation.
Therefore, the equatorial ejecta have lower electron fractions ($\ye \approx 0.2 - 0.25$).
Furthermore, the energy deposition by neutrinos is reduced, so the amount of mass ejected in this direction decreases continuously as the oscillations of the massive \ac{NS} die down.

\subsection{Correlation of ejecta masses to EOS properties}

\begin{figure}
  \centering
  \includegraphics{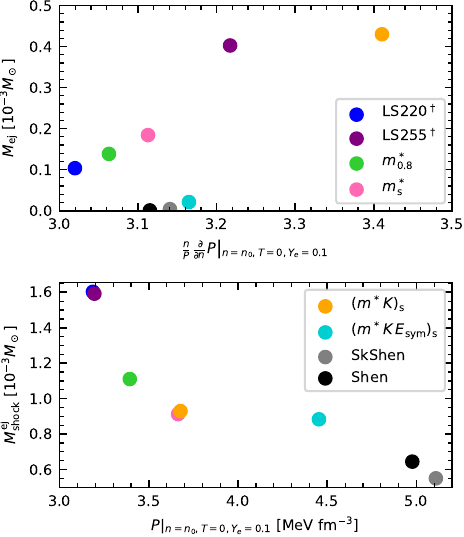}
  \caption{%
    Masses of the tidal and shock-heated ejecta components as a function of the dimensionless slope of the pressure and the pressure at saturation density, respectively.
    \label{fig:ejecta_correlations}}
\end{figure}

The masses of the ejecta components at $t-t_{\text{merg}} = \unit[40]{ms}$ are given in \cref{tab:overview} for all simulations.
In the following analysis, we will focus on the correlation of the dynamical ejecta masses with the properties of the employed \acp{EOS}.
Since the mechanism of ejection is entirely different for the tidal and shock-heated dynamical ejecta components, it is not surprising that they depend differently on the properties of the EOS.
This is visible in \cref{fig:ejecta_correlations}, which shows the masses of the tidal and shock-heated ejecta versus the dimensionless slope of the pressure and the absolute pressure at saturation density of the corresponding EOS (as defined in \cref{sec:eos_tables}), respectively.
\begin{figure*}
  \centering
  \includegraphics{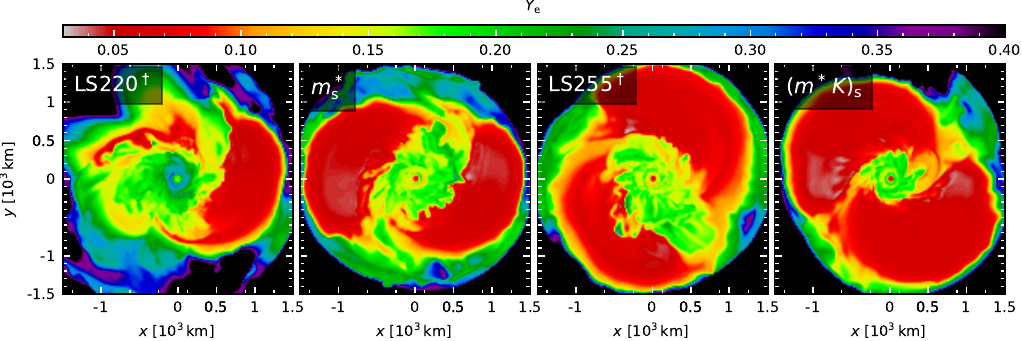}
  \caption{%
    Distribution of $\ye$ in the $xy$ plane for the models \LSm, \msS, \LSl, and \msKS \unit[15]{ms} after the merger.
    The tidal ejecta are visible as large, smooth areas with low $\ye$.
    The dimensionless slope of the pressure and accordingly the size of the tidal arms increases from the left to the right panels.
    \label{fig:tidal_arms}}
\end{figure*}

It is well-known in the literature that the shock-heated ejecta mass is correlated with the \ac{EOS} stiffness \cite[see, \eg,][]{Dietrich2017a, Radice2018a, Nedora2021c, Henkel2023a}.
This is often quantified by properties of cold \acp{NS} (like the reduced tidal deformability), which depend on the full range of densities present in \acp{NS} and thus also on the slope of the pressure at saturation density.
We find that the shock-heated ejecta mass is correlated with pressure at saturation density and entirely unaffected by the slope of the pressure at saturation density.
Specifically, we find that pairs of models that only differ in the incompressibility (\LSm and \LSh or \msS and \msKS) exhibit almost the same amount of shock-heated ejecta.
This points towards the shock-heated ejection mechanism being most sensitive to the \ac{EOS} close to the saturation density and entirely independent of the high-density part of the \ac{EOS}.

The mass of the tidal ejecta varies between $10^{-6} - 4.3 \times 10^{-3} M_\odot$.
Even though it is relatively small in comparison to the other ejecta components, it plays a significant role because it can produce actinides and fissioning isotopes due to its very low electron fraction.
This is important for galactic chemical evolution and the kilonova light curve, which might show signatures of fission reactions if very heavy elements are produced.
Furthermore, the tidal ejecta component is expected to be larger in the merger of asymmetric binaries.
In the models \msKES, SkShen, and Shen, the tidal ejecta are caught by shocks early on, so almost no material with $\ye < 0.1$ remains.
While this implies that almost no tidal ejecta component exists in these models by our definition, it does not mean that no matter is ejected by tidal torques.
For the other models, the tidal ejecta mass shows a correlation with the dimensionless slope of the pressure and is significantly larger for \LSh and \msKS compared to the other models.
\Cref{fig:tidal_arms} shows the distribution of the electron fraction in the equatorial plane for \LSm, \msS, \LSh, and \msKS.
The tidal arms are visible as the low electron-fraction (\ie, red) regions, which are larger for models with larger slopes of the pressure.

\begin{figure}
  \centering
  \includegraphics{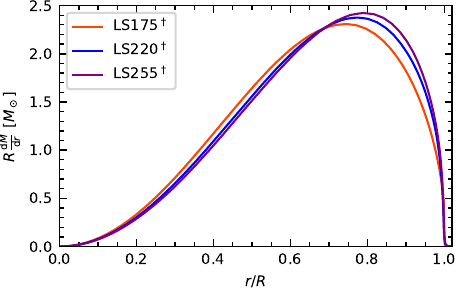}
  \caption{%
    Spatial mass distribution $\frac{\mathrm{d} M}{\mathrm{d} r} = 4 \pi \rho r^2$ versus the radius for \LSl, \LSm, and \LSh.
    The radius is normalized to the radius of the \ac{NS}.
    The total mass of the \ac{NS} is given by the area under the curve.
    \label{fig:rho_TOV}}
\end{figure}
The fact that a larger slope of the pressure influences the tidal ejecta mass might be related to the fact that it affects the distribution of the mass inside the \ac{NS}.
For a large slope of the pressure with respect to the density, the central density in the \ac{NS} is reduced but at the same time, the density in the outer layers is increased.
This effect can be seen by comparing the mass distribution in the \acp{NS} in the initial data of our simulations.
\Cref{fig:rho_TOV} shows the spatial mass distribution $\frac{\mathrm{d} M}{\mathrm{d} r} = 4 \pi \rho r^2$ versus the radius for the models \LSl, \LSm, and \LSh.
For the sake of comparison, the radius is normalized to the outermost radius of the \ac{NS} $R$.
Since there is more mass located in the outer layers of the \acp{NS}, the ejection of matter by tidal forces is more efficient.


\section{Gravitational wave emission}
\label{sec:gravitational_wave_emission}

\begin{figure*}
  \centering
  \includegraphics{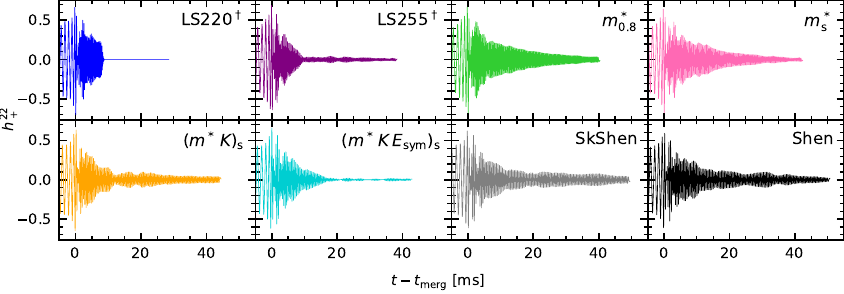}
  \caption{%
    Waveforms of the \ac{GW} signals for all simulations except
    \LSl. \label{fig:waveforms}}
\end{figure*}

We extract the gravitational waveform as outlined in \cref{sec:postprocessing}.
\mbox{\Cref{fig:waveforms}} shows the $+$ mode of the \ac{GW} strain for all simulations except \LSl, since no significant post-merger signal is produced after the prompt collapse.
We only show the $l=2,m=2$ mode as it is by far the most dominant mode.
It shows the low-frequency pre-merger phase ($t < 0$), as well as the post-merger phase ($t > 0$).
The latter consists of two periods.
A transition period ($t \lesssim \unit[3]{ms}$), during which the system readjusts from the inspiral of two \acp{NS} to one rotating and oscillating massive \ac{NS}, and the longer ring-down phase that follows, during which the amplitude gradually decreases \ceg{Baiotti2019a}.
In \LSm, the \ac{GW} signal stops abruptly after the collapse of the hypermassive \ac{NS}.
In \LSh, \msKES, and \msKS, the \ac{GW} amplitude drops within the first \unit[10-15]{ms} while it decreases more slowly within \unit[20-30]{ms} in \msm, \msS, SkShen, and Shen.
This decrease in the amplitude is directly linked to the $m=2$ deformation of the massive \ac{NS} (see \cref{fig:xy_decomp}).
As described in \cref{sec:disk}, the $m=2$ deformation decreases much faster in the simulations with higher pressure slopes.
Future detections of post-merger \acp{GW} could therefore be used to constrain the slope of the pressure at high densities.

\begin{figure*}
  \centering
  \includegraphics{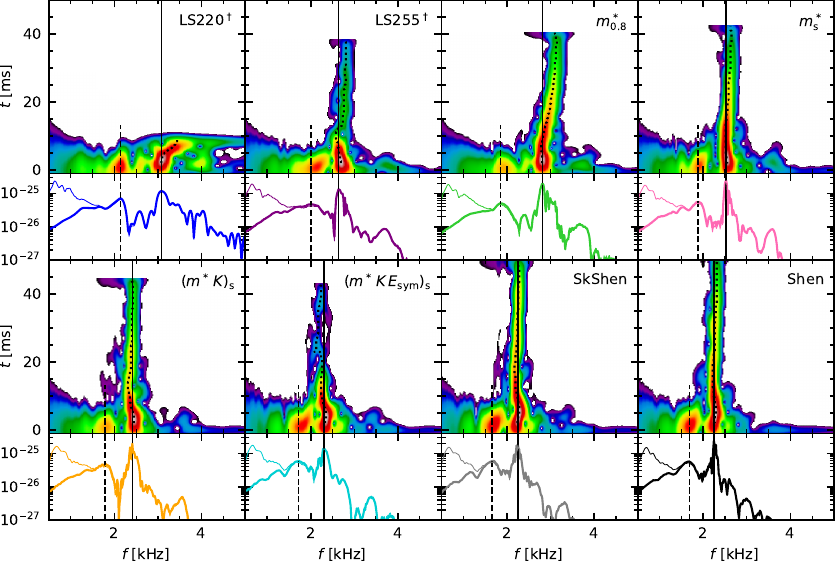}
  \caption{%
    Post-merger \ac{GW} Fourier spectra and corresponding time-frequency spectrograms for all simulations except \LSl.
    The thin lines in the Fourier spectra show the full spectrum, while the thick lines are produced by excluding the inspiral \ac{GW} signal.
    Dashed and solid vertical lines mark the position of the $f_1$ and $f_2$ peak frequencies extracted from the time-independent Fourier spectra, and dotted lines show the time-dependent peak frequency extracted from the time-frequency map.
    \label{fig:spectrograms}}
\end{figure*}

The post-merger \ac{GW} time-frequency spectrograms (upper panels) and the corresponding Fourier spectra (lower panels) are shown in \cref{fig:spectrograms}.
The thin lines in the Fourier spectra represent the full spectrum, while the thick lines are produced by excluding the inspiral \ac{GW} signal.
Initially, several frequencies are present but they decay within approximately 5 milliseconds \citep{Takami2015a, Rezzolla2016a}.
Afterward, the spectrum is dominated by a single frequency, often called $f_2$.

This frequency has been identified and studied in many previous works \citep{Stergioulas2011a, Bauswein2012a, Bauswein2012b, Hotokezaka2013b, Takami2014a, Takami2015a, Rezzolla2016a, Fields2023a} and is attributed to the quadrupole mode of the \ac{NS} \citep{Bauswein2012a, Bauswein2012b}.
The solid vertical lines show the maximum of the Fourier spectra, while the dotted lines follow the time-dependent maximum frequency.
The following effects can be seen in the time-frequency spectrograms:
\begin{itemize}
\item In \LSm, the post-merger signal shows a ``chirp-like'' behavior (the peak frequency rises quickly) shortly before the remnant collapses to a \ac{BH}.
  This is because the rotational velocity of the massive \ac{NS} increases significantly as it collapses.
\item
  In the first \unit[10]{ms} of the post-merger \ac{GW} emission, the $f_2$ frequency can vary slightly.
  This is especially visible for the stiffest \acp{EOS}: \msKS, \msKES, SkShen, and Shen.
  For those, the $f_2$ frequency increases until $t \approx \unit[5]{ms}$ and subsequently decreases again until $t \approx \unit[10]{ms}$.
  A similar effect is described by \citet{Rezzolla2016a}.
  They determine the frequency during the transient phase separately from the ring-down phase and label it $f_{2, i}$.
\item
  After the initial transient phase, a continuous shift of the $f_2$ emission towards higher frequencies is visible for \msm and \msS.
  A similar but weaker increase is visible for \msS, SkShen, and Shen.
  In these models, the gravitational wave amplitude stays high for an extended period.
  The \acp{GW} carry away angular momentum, the \ac{NS} contracts leading to a larger $f_2$ \citep{Maione2017a}.
\end{itemize}
Similar trends can be seen in the spectrograms in \citet{Rezzolla2016a, Dietrich2017a, Maione2017a}.

The second most prominent peak is the so-called $f_1$ peak (sometimes also called $f_-$), which always lies at lower frequencies than the $f_2$ peak and disappears after $\sim$ \unit[5]{ms} \citep{Stergioulas2011a, Takami2014a, Takami2015a, Rezzolla2016a}.
Its origin has been attributed to the interaction of the $f_2$ and the quasi-radial $f_0$ mode (named $f_{2-0}$) \citep{Stergioulas2011a} as well as the orbital motion of antipodal bulges rotating around the central remnant with a slower frequency (thus called $f_{\text{spiral}}$) \citep{Bauswein2015a}.
Depending on the remnant compactness, either one or both frequencies might be present \citep{Bauswein2015a, Bauswein2016a, Rezzolla2016a, Maione2017a, Kiuchi2020a}.
We define $f_1$ independently of its origin as the second highest peak with a frequency at least \unit[400]{Hz} below the $f_2$ peak.
The extracted $f_1$ peaks are marked by a dashed vertical line in \cref{fig:spectrograms}.

The $f_1$ and $f_2$ frequencies for all simulations are listed in \cref{tab:overview}.
They span from $f_2 = \unit[2.24]{kHz}$ for Shen to $f_2 = \unit[3.10]{kHz}$ for \LSm and $f_1 = \unit[1.66]{kHz}$ for SkShen to $f_1 = \unit[2.14]{kHz}$ for \LSm where softer \acp{EOS} generally produce larger frequencies.
Decreasing the effective mass and increasing the incompressibility and symmetry energy lowers the pressure in the center of the \ac{NS}.
This decreases the density, which in turn reduces $f_2$ \citep{Bauswein2012a, Bauswein2012b}.
By changing the nuclear matter properties to the values of the Shen \ac{EOS} (\ie, the progression \msS, \msKS, \msKES, SkShen), both peak frequencies approach those of the Shen simulation.

The Fourier spectra and the time-frequency spectrograms of models Shen and SkShen are very similar.
Especially the position and width of the $f_1$ and $f_2$ peaks as well as the amplitude and time dependence of the spectrogram match almost perfectly.
This implies, that the \ac{EOS}-impact on the post-merger \ac{GW} emission is well described by the nuclear matter properties of the \ac{EOS}, while the details of the microphysics and models (\ie, relativistic mean field and Skyrme density functionals) only play a minor role.
It might thus be possible to directly constrain the properties of nuclear matter with future detections of the post-merger \ac{GW} spectra.
Furthermore, this implies that the post-merger \ac{GW} emission is mostly sensitive to the \ac{EOS} around saturation density, since this is the region where SkShen and Shen match.

\begin{figure}
  \centering
  \includegraphics{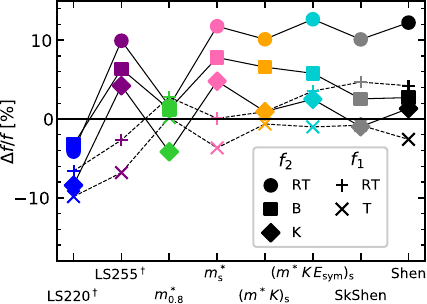}
  \caption{%
    Comparison of the post-merger \ac{GW} spectrum with various fit formulae from the literature.
    Included are fit formulae for $f_1$ and $f_2$ from \citet{Rezzolla2016a}, Eqs. (20) and (23) (labeled RT),
    \citet{Kiuchi2020a}, Eq. (5.4) (labeled K),
    \citet{Bauswein2016a}, Eq. (2) (labeled B),
    and \citet{Takami2015a}, Eq. (25) (labeled T).
    Shown are the relative deviations of the fits to the measured frequencies $\Delta f = f_{\text{NR}} - f_{\text{fit}}$.
    Solid symbols connected with solid lines represent fits for $f_2$, while crosses connected with dashed lines represent fits for $f_1$.
    \label{fig:univ_rel}}
\end{figure}
Many works have found that the $f_2$ frequency is correlated with the properties of the \ac{EOS}.
\Citet{Takami2015a, Rezzolla2016a, Bauswein2016a, Kiuchi2020a} have provided fit formulae relating $f_1$ and $f_2$ with various properties of cold isolated \acp{NS}.
We compare their fit formulae with our results.
\Cref{fig:univ_rel} shows the residuals of the fits to our extracted frequencies.

All fit formulae predict the correct frequencies within an uncertainty of 5$-$10\%.
However, some trends in the difference can be identified.
Fit formulae for $f_2$ (solid symbols connected with solid lines) underestimate the value for $f_2$ in \LSm and overestimate it in \LSh, \msS, \msKS, \msKES, SkShen, and Shen.
The fit from \cite{Kiuchi2020a} performs better, but also uses the most fit parameters.
The fits for $f_1$ do relatively well for the models with $m^*/m < 1$ but overestimate the frequency for \LSh and \LSm.
Our set of simulations is too small to make quantitative predictions for the peak frequencies.
However, we plan to increase the number of models in future works.
Therefore, we will potentially be able to enhance the accuracy of universal relations in the future by incorporating additional parameters describing the \ac{EOS} into the fits.


\section{Conclusions}
\label{sec:conclusions}

In this work, we systematically investigated the influence of the nuclear \ac{EOS} on \ac{BNS} mergers.
To this end, we performed simulations of merging equal mass \ac{NS} binaries while individually varying the nuclear matter properties in the employed \acp{EOS} from the fiducial model \LSm to those of the Shen \ac{EOS}.
This study focused on the influence of the nucleon effective mass and the incompressibility parameter.
We provide a comprehensive analysis of the remnant and disc dynamics, the post-merger \ac{GW} spectrum, as well as the properties of the dynamical and early disk ejecta.
The key findings can be summarized as follows:
\begin{itemize}
\item
  The incompressibility governs the slope of the cold pressure at saturation density.
  Therefore, it has a large impact when the density is much larger than saturation density, \ie, in the core of the merger remnant, which typically reaches three to five times saturation density.
  This has implications for the emission of gravitational waves, as well as for the fate of the remnant.
  Increasing the incompressibility decreases the compactness of the remnant core and therefore lowers the dominant post-merger \ac{GW} frequency.
  For our choice of initial conditions, varying the incompressibility $K$ within theoretical uncertainties results in a prompt collapse for low $K$ and a \ac{NS} remnant that is stable for the duration of the simulation for high $K$.
  In the latter case, the steep rise of the pressure halts the contraction of the newly formed remnant shortly after collapse and therefore leads to less oscillations in the remnant compared to other non-collapsing simulations.
\item
  The effective nucleon mass is important for both the pressure at zero temperature as well as thermal contributions in our EOS models.
  Lowering the effective mass increases the cold pressure at all densities inside the \ac{NS}.
  Similar to the model with increased incompressibility, this results in a less compact remnant and a reduction of the peak frequency in the post-merger \ac{GW} spectrum.
  However, a significant difference is that the pressure density dependence is less steep leading to a longer period of contraction and oscillations as imprinted in the post-merger \ac{GW} amplitude.
  Another important aspect of the effective mass is its influence on the
  thermal pressure.
  Decreasing the effective mass results in a larger thermal index.
  At the same time, it also reduces shock heating during and after the plunge
  because the remnant is less compact.
  Thus, the total thermal pressure is reduced for lower effective masses, even
  though the thermal index is higher.
\item
  The dynamical ejecta mass is correlated with the effective mass and the incompressibility.
  On the one hand, larger incompressibilities and lower effective masses reduce the compactness of the initial \ac{NS} and therefore increase the amount of tidal ejecta.
  This effect is especially strong for higher incompressibilities.
  The mass of the shock-heated ejecta, on the other hand, are mostly correlated by the effective mass.
  Mass ejection in the early disk phase is more complex because multiple effects, such as remnant deformation, oscillations, and neutrino emission play a role.
\item
  We matched all nuclear matter properties to those of Shen \ac{EOS}.
  The resulting SkShen \ac{EOS} is mostly similar to the Shen \ac{EOS} except for larger pressures at high densities and a slightly lower thermal index.
  In the corresponding simulations, the evolution of the remnant and the accretion disk, as well as the amounts of ejected mass are relatively similar, and the post-merger \ac{GW} spectrum of the two models is remarkably similar.
  Therefore, we conclude that nuclear matter properties are a useful measure to quantify EOS effects in BNS mergers.
\end{itemize}

Nuclear matter properties are a promising way to quantify \ac{EOS} effects in \ac{BNS} mergers that is independent of the microphysical framework used.
Furthermore, they can be constrained by nuclear physics experiments and theoretical calculations, enabling the combination of results from nuclear physics, astrophysics, and multi-messenger astronomy.
However, more studies investigating the role of \ac{EOS} properties in compact object mergers are needed in the future.


\section*{Acknowledgements}

We thank Sho Fujibayashi, Bruno Giacomazzo, Wolfgang Kastaun, Takami Kuroda, and Martin Obergaulinger for useful discussions.
This work was supported by the Deutsche Forschungsgemeinschaft (DFG, German Research Foundation) -- Project-ID 279384907 -- SFB 1245, the State of Hessen within the Research Cluster ELEMENTS (Project ID 500/10.006), and the European Research Council under grant EUROPIUM-667912.
FMG acknowledges funding from the Fondazione CARITRO, program Bando post-doc 2021, project number 11745.
The simulations were performed on HPE Apollo HAWK at the High-Performance Computing Center Stuttgart (HLRS) under the grant microBNS.

\section*{Data Availability}

The data underlying this article will be shared on reasonable request to the corresponding authors.

\bibliographystyle{mnras}
\bibliography{ms}

\begin{thebibliography}{}
\makeatletter
\relax
\def\mn@urlcharsother{\let\do\@makeother \do\$\do\&\do\#\do\^\do\_\do\%\do\~}
\def\mn@doi{\begingroup\mn@urlcharsother \@ifnextchar [ {\mn@doi@}
  {\mn@doi@[]}}
\def\mn@doi@[#1]#2{\def\@tempa{#1}\ifx\@tempa\@empty \href
  {http://dx.doi.org/#2} {doi:#2}\else \href {http://dx.doi.org/#2} {#1}\fi
  \endgroup}
\def\mn@eprint#1#2{\mn@eprint@#1:#2::\@nil}
\def\mn@eprint@arXiv#1{\href {http://arxiv.org/abs/#1} {{\tt arXiv:#1}}}
\def\mn@eprint@dblp#1{\href {http://dblp.uni-trier.de/rec/bibtex/#1.xml}
  {dblp:#1}}
\def\mn@eprint@#1:#2:#3:#4\@nil{\def\@tempa {#1}\def\@tempb {#2}\def\@tempc
  {#3}\ifx \@tempc \@empty \let \@tempc \@tempb \let \@tempb \@tempa \fi \ifx
  \@tempb \@empty \def\@tempb {arXiv}\fi \@ifundefined
  {mn@eprint@\@tempb}{\@tempb:\@tempc}{\expandafter \expandafter \csname
  mn@eprint@\@tempb\endcsname \expandafter{\@tempc}}}

\bibitem[\protect\citeauthoryear{Abbott et~al.,}{Abbott
  et~al.}{2017a}]{Abbott2017a}
Abbott B.~P.,  et~al., 2017a, \mn@doi [Phys. Rev. Lett.]
  {10.1103/PhysRevLett.119.161101}, 119, 161101

\bibitem[\protect\citeauthoryear{Abbott et~al.,}{Abbott
  et~al.}{2017b}]{Abbott2017b}
Abbott B.~P.,  et~al., 2017b, \mn@doi [ApJ] {10.3847/2041-8213/aa91c9}, 848,
  L12

\bibitem[\protect\citeauthoryear{Abbott et~al.,}{Abbott
  et~al.}{2018}]{Abbott2018a}
Abbott B.~P.,  et~al., 2018, \mn@doi [Phys. Rev. Lett.]
  {10.1103/PHYSREVLETT.121.161101}, 121, 161101

\bibitem[\protect\citeauthoryear{Abbott et~al.,}{Abbott
  et~al.}{2019}]{Abbott2019a}
Abbott B.~P.,  et~al., 2019, \mn@doi [Phys. Rev. X]
  {10.1103/PhysRevX.9.011001}, 9, 011001

\bibitem[\protect\citeauthoryear{Adhikari et~al.,}{Adhikari
  et~al.}{2021}]{Adhikari2021a}
Adhikari D.,  et~al., 2021, \mn@doi [Phys. Rev. Lett.]
  {10.1103/PhysRevLett.126.172502}, 126, 172502

\bibitem[\protect\citeauthoryear{Adhikari et~al.,}{Adhikari
  et~al.}{2022}]{Adhikari2022a}
Adhikari D.,  et~al., 2022, \mn@doi [Phys. Rev. Lett.]
  {10.1103/PhysRevLett.129.042501}, 129, 042501

\bibitem[\protect\citeauthoryear{Andersen, Zha, Schneider, Betranhandy, Couch
  \& O'Connor}{Andersen et~al.}{2021}]{Andersen2021a}
Andersen O.~E.,  Zha S.,  Schneider A. d.~S.,  Betranhandy A.,  Couch S.~M.,
  O'Connor E.~P.,  2021, \mn@doi [ApJ] {10.3847/1538-4357/ac294c}, 923, 201

\bibitem[\protect\citeauthoryear{Andreoni et~al.,}{Andreoni
  et~al.}{2017}]{Andreoni2017a}
Andreoni I.,  et~al., 2017, \mn@doi [Publ. Astron. Soc. Australia]
  {10.1017/PASA.2017.65}, 34

\bibitem[\protect\citeauthoryear{Antoniadis et~al.,}{Antoniadis
  et~al.}{2013}]{Antoniadis2013a}
Antoniadis J.,  et~al., 2013, \mn@doi [Science] {10.1126/science.1233232}, 340,
  1233232

\bibitem[\protect\citeauthoryear{Arcavi et~al.,}{Arcavi
  et~al.}{2017}]{Arcavi2017a}
Arcavi I.,  et~al., 2017, \mn@doi [Nature] {10.1038/NATURE24291}, 551, 64

\bibitem[\protect\citeauthoryear{Baiotti}{Baiotti}{2019}]{Baiotti2019a}
Baiotti L.,  2019, \mn@doi [Prog. Part. Nucl. Phys.]
  {10.1016/j.ppnp.2019.103714}, 109, 103714

\bibitem[\protect\citeauthoryear{Baumgarte \& Shapiro}{Baumgarte \&
  Shapiro}{1998}]{Baumgarte1998a}
Baumgarte T.~W.,  Shapiro S.~L.,  1998, \mn@doi [Phys. Rev. D]
  {10.1103/PhysRevD.59.024007}, 59, 024007

\bibitem[\protect\citeauthoryear{Baumgarte \& Shapiro}{Baumgarte \&
  Shapiro}{2021}]{Baumgarte2021a}
Baumgarte T.~W.,  Shapiro S.~L.,  2021, Numerical {{Relativity}}: {{Starting}}
  from {{Scratch}}, 1 edn.
{Cambridge University Press}, {Cambridge}

\bibitem[\protect\citeauthoryear{Bauswein \& Janka}{Bauswein \&
  Janka}{2012}]{Bauswein2012a}
Bauswein A.,  Janka H.-T.,  2012, \mn@doi [Phys. Rev. Lett.]
  {10.1103/PhysRevLett.108.011101}, 108, 011101

\bibitem[\protect\citeauthoryear{Bauswein \& Stergioulas}{Bauswein \&
  Stergioulas}{2015}]{Bauswein2015a}
Bauswein A.,  Stergioulas N.,  2015, \mn@doi [Phys. Rev. D]
  {10.1103/PhysRevD.91.124056}, 91, 124056

\bibitem[\protect\citeauthoryear{Bauswein, Janka  \& Oechslin}{Bauswein
  et~al.}{2010}]{Bauswein2010a}
Bauswein A.,  Janka H.-T.,   Oechslin R.,  2010, \mn@doi [Phys. Rev. D]
  {10.1103/PhysRevD.82.084043}, 82, 084043

\bibitem[\protect\citeauthoryear{Bauswein, Janka, Hebeler  \& Schwenk}{Bauswein
  et~al.}{2012}]{Bauswein2012b}
Bauswein A.,  Janka H.-T.,  Hebeler K.,   Schwenk A.,  2012, \mn@doi [Phys.
  Rev. D] {10.1103/PhysRevD.86.063001}, 86, 063001

\bibitem[\protect\citeauthoryear{Bauswein, Goriely  \& Janka}{Bauswein
  et~al.}{2013}]{Bauswein2013b}
Bauswein A.,  Goriely S.,   Janka H.-T.,  2013, \mn@doi [ApJ]
  {10.1088/0004-637X/773/1/78}, 773, 78

\bibitem[\protect\citeauthoryear{Bauswein, Stergioulas  \& Janka}{Bauswein
  et~al.}{2016}]{Bauswein2016a}
Bauswein A.,  Stergioulas N.,   Janka H.~T.,  2016, \mn@doi [Eur. Phys. J. A]
  {10.1140/epja/i2016-16056-7}, 52

\bibitem[\protect\citeauthoryear{Bauswein, Blacker, Lioutas, Soultanis, Vijayan
   \& Stergioulas}{Bauswein et~al.}{2021}]{Bauswein2021a}
Bauswein A.,  Blacker S.,  Lioutas G.,  Soultanis T.,  Vijayan V.,
  Stergioulas N.,  2021, \mn@doi [Phys. Rev. D] {10.1103/PhysRevD.103.123004},
  103, 123004

\bibitem[\protect\citeauthoryear{Berger \& Colella}{Berger \&
  Colella}{1989}]{Berger1989a}
Berger M.~J.,  Colella P.,  1989, \mn@doi [J. Comput. Phys.]
  {10.1016/0021-9991(89)90035-1}, 82, 64

\bibitem[\protect\citeauthoryear{Berger \& Oliger}{Berger \&
  Oliger}{1984}]{Berger1984a}
Berger M.~J.,  Oliger J.,  1984, \mn@doi [J. Comput. Phys.]
  {10.1016/0021-9991(84)90073-1}, 53, 484

\bibitem[\protect\citeauthoryear{Bernuzzi}{Bernuzzi}{2020}]{Bernuzzi2020a}
Bernuzzi S.,  2020, \mn@doi [General Relativ. Gravitation]
  {10.1007/s10714-020-02752-5}, 52, 108

\bibitem[\protect\citeauthoryear{Bernuzzi, Dietrich, Tichy  \&
  Br{\"u}gmann}{Bernuzzi et~al.}{2014}]{Bernuzzi2014a}
Bernuzzi S.,  Dietrich T.,  Tichy W.,   Br{\"u}gmann B.,  2014, \mn@doi [Phys.
  Rev. D] {10.1103/PhysRevD.89.104021}, 89, 104021

\bibitem[\protect\citeauthoryear{Birkhan et~al.,}{Birkhan
  et~al.}{2017}]{Birkhan2017a}
Birkhan J.,  et~al., 2017, \mn@doi [Phys. Rev. Lett.]
  {10.1103/PhysRevLett.118.252501}, 118, 252501

\bibitem[\protect\citeauthoryear{Bovard, Martin, Guercilena, Arcones, Rezzolla
  \& Korobkin}{Bovard et~al.}{2017}]{Bovard2017b}
Bovard L.,  Martin D.,  Guercilena F.,  Arcones A.,  Rezzolla L.,   Korobkin
  O.,  2017, \mn@doi [Phys. Rev. D] {10.1103/PhysRevD.96.124005}, 96, 124005

\bibitem[\protect\citeauthoryear{Brown}{Brown}{2009}]{Brown2009a}
Brown J.~D.,  2009, \mn@doi [Phys. Rev. D] {10.1103/PhysRevD.79.104029}, 79,
  104029

\bibitem[\protect\citeauthoryear{Brown, Diener, Sarbach, Schnetter  \&
  Tiglio}{Brown et~al.}{2009}]{Brown2009b}
Brown D.,  Diener P.,  Sarbach O.,  Schnetter E.,   Tiglio M.,  2009, \mn@doi
  [Phys. Rev. D] {10.1103/PhysRevD.79.044023}, 79, 044023

\bibitem[\protect\citeauthoryear{Carbone \& Schwenk}{Carbone \&
  Schwenk}{2019}]{Carbone2019a}
Carbone A.,  Schwenk A.,  2019, \mn@doi [Phys. Rev. C]
  {10.1103/PhysRevC.100.025805}, 100, 025805

\bibitem[\protect\citeauthoryear{Combi \& Siegel}{Combi \&
  Siegel}{2023}]{Combi2023a}
Combi L.,  Siegel D.~M.,  2023, \mn@doi [ApJ] {10.3847/1538-4357/acac29}, 944,
  28

\bibitem[\protect\citeauthoryear{Constantinou, Muccioli, Prakash  \&
  Lattimer}{Constantinou et~al.}{2015}]{Constantinou2015a}
Constantinou C.,  Muccioli B.,  Prakash M.,   Lattimer J.~M.,  2015, \mn@doi
  [Phys. Rev. C] {10.1103/PhysRevC.92.025801}, 92, 025801

\bibitem[\protect\citeauthoryear{C{\^o}t{\'e} et~al.,}{C{\^o}t{\'e}
  et~al.}{2019}]{Cote2019a}
C{\^o}t{\'e} B.,  et~al., 2019, \mn@doi [ApJ] {10.3847/1538-4357/AB10DB}, 875,
  106

\bibitem[\protect\citeauthoryear{Coulter et~al.,}{Coulter
  et~al.}{2017}]{Coulter2017a}
Coulter D.~A.,  et~al., 2017, \mn@doi [Science] {10.1126/SCIENCE.AAP9811}, 358,
  1556

\bibitem[\protect\citeauthoryear{Cowperthwaite et~al.,}{Cowperthwaite
  et~al.}{2017}]{Cowperthwaite2017a}
Cowperthwaite P.~S.,  et~al., 2017, \mn@doi [ApJ] {10.3847/2041-8213/AA8FC7},
  848, L17

\bibitem[\protect\citeauthoryear{Danielewicz, Lacey  \& Lynch}{Danielewicz
  et~al.}{2002}]{Danielewicz2002a}
Danielewicz P.,  Lacey R.,   Lynch W.~G.,  2002, \mn@doi [Science]
  {10.1126/science.1078070}, 298, 1592

\bibitem[\protect\citeauthoryear{De, Finstad, Lattimer, Brown, Berger  \&
  Biwer}{De et~al.}{2018}]{De2018a}
De S.,  Finstad D.,  Lattimer J.~M.,  Brown D.~A.,  Berger E.,   Biwer C.~M.,
  2018, \mn@doi [Phys. Rev. Lett.] {10.1103/PhysRevLett.121.091102}, 121,
  091102

\bibitem[\protect\citeauthoryear{Dean, Fern{\'a}ndez  \& Metzger}{Dean
  et~al.}{2021}]{Dean2021a}
Dean C.,  Fern{\'a}ndez R.,   Metzger B.~D.,  2021, \mn@doi [ApJ]
  {10.3847/1538-4357/ac1f20}, 921, 161

\bibitem[\protect\citeauthoryear{Demorest, Pennucci, Ransom, Roberts  \&
  Hessels}{Demorest et~al.}{2010}]{Demorest2010a}
Demorest P.~B.,  Pennucci T.,  Ransom S.~M.,  Roberts M. S.~E.,   Hessels J.
  W.~T.,  2010, \mn@doi [Nature] {10.1038/nature09466}, 467, 1081

\bibitem[\protect\citeauthoryear{Dessart, Ott, Burrows, Rosswog  \&
  Livne}{Dessart et~al.}{2009}]{Dessart2009a}
Dessart L.,  Ott C.~D.,  Burrows A.,  Rosswog S.,   Livne E.,  2009, \mn@doi
  [ApJ] {10.1088/0004-637X/690/2/1681}, 690, 1681

\bibitem[\protect\citeauthoryear{D{\'i}az et~al.,}{D{\'i}az
  et~al.}{2017}]{Diaz2017a}
D{\'i}az M.~C.,  et~al., 2017, \mn@doi [ApJ] {10.3847/2041-8213/AA9060}, 848,
  L29

\bibitem[\protect\citeauthoryear{Dietrich \& Ujevic}{Dietrich \&
  Ujevic}{2017}]{Dietrich2017b}
Dietrich T.,  Ujevic M.,  2017, \mn@doi [Classical Quantum Gravity]
  {10.1088/1361-6382/aa6bb0}, 34, 105014

\bibitem[\protect\citeauthoryear{Dietrich, Ujevic, Tichy, Bernuzzi  \&
  Br{\"u}gmann}{Dietrich et~al.}{2017}]{Dietrich2017a}
Dietrich T.,  Ujevic M.,  Tichy W.,  Bernuzzi S.,   Br{\"u}gmann B.,  2017,
  \mn@doi [Phys. Rev. D] {10.1103/PhysRevD.95.024029}, 95, 024029

\bibitem[\protect\citeauthoryear{Dietrich, Coughlin, Pang, Bulla, Heinzel,
  Issa, Tews  \& Antier}{Dietrich et~al.}{2020}]{Dietrich2020a}
Dietrich T.,  Coughlin M.~W.,  Pang P. T.~H.,  Bulla M.,  Heinzel J.,  Issa L.,
   Tews I.,   Antier S.,  2020, \mn@doi [Science] {10.1126/science.abb4317},
  370, 1450

\bibitem[\protect\citeauthoryear{Drischler, Hebeler  \& Schwenk}{Drischler
  et~al.}{2016}]{Drischler2016a}
Drischler C.,  Hebeler K.,   Schwenk A.,  2016, \mn@doi [Phys. Rev. C]
  {10.1103/PhysRevC.93.054314}, 93, 054314

\bibitem[\protect\citeauthoryear{Drischler, Hebeler  \& Schwenk}{Drischler
  et~al.}{2019}]{Drischler2019a}
Drischler C.,  Hebeler K.,   Schwenk A.,  2019, \mn@doi [Phys. Rev. Lett.]
  {10.1103/PhysRevLett.122.042501}, 122, 042501

\bibitem[\protect\citeauthoryear{Drischler, Furnstahl, Melendez  \&
  Phillips}{Drischler et~al.}{2020}]{Drischler2020a}
Drischler C.,  Furnstahl R.~J.,  Melendez J.~A.,   Phillips D.~R.,  2020,
  \mn@doi [Phys. Rev. Lett.] {10.1103/PhysRevLett.125.202702}, 125, 202702

\bibitem[\protect\citeauthoryear{Drout et~al.,}{Drout
  et~al.}{2017}]{Drout2017a}
Drout M.~R.,  et~al., 2017, \mn@doi [Science] {10.1126/science.aaq0049}, 358,
  1570

\bibitem[\protect\citeauthoryear{East, Paschalidis  \& Pretorius}{East
  et~al.}{2016a}]{East2016a}
East W.~E.,  Paschalidis V.,   Pretorius F.,  2016a, \mn@doi [Classical Quantum
  Gravity] {10.1088/0264-9381/33/24/244004}, 33, 244004

\bibitem[\protect\citeauthoryear{East, Paschalidis, Pretorius  \& Shapiro}{East
  et~al.}{2016b}]{East2016b}
East W.~E.,  Paschalidis V.,  Pretorius F.,   Shapiro S.~L.,  2016b, \mn@doi
  [Phys. Rev. D] {10.1103/PHYSREVD.93.024011}, 93, 024011

\bibitem[\protect\citeauthoryear{Evans et~al.,}{Evans
  et~al.}{2017}]{Evans2017a}
Evans P.~A.,  et~al., 2017, \mn@doi [Science] {10.1126/SCIENCE.AAP9580}, 358,
  1565

\bibitem[\protect\citeauthoryear{Fahlman \& Fern{\'a}ndez}{Fahlman \&
  Fern{\'a}ndez}{2022}]{Fahlman2022a}
Fahlman S.,  Fern{\'a}ndez R.,  2022, \mn@doi [MNRAS] {10.1093/mnras/stac948},
  513, 2689

\bibitem[\protect\citeauthoryear{Fern{\'a}ndez \& Metzger}{Fern{\'a}ndez \&
  Metzger}{2013}]{Fernandez2013a}
Fern{\'a}ndez R.,  Metzger B.~D.,  2013, \mn@doi [MNRAS]
  {10.1093/mnras/stt1312}, 435, 502

\bibitem[\protect\citeauthoryear{Fields, Prakash, Breschi, Radice, Bernuzzi  \&
  Schneider}{Fields et~al.}{2023}]{Fields2023a}
Fields J.,  Prakash A.,  Breschi M.,  Radice D.,  Bernuzzi S.,   Schneider A.
  d.~S.,  2023, \mn@doi [ApJ] {10.3847/2041-8213/ace5b2}, 952, L36

\bibitem[\protect\citeauthoryear{Fonseca et~al.,}{Fonseca
  et~al.}{2021}]{Fonseca2021a}
Fonseca E.,  et~al., 2021, \mn@doi [ApJ] {10.3847/2041-8213/AC03B8}, 915, L12

\bibitem[\protect\citeauthoryear{Foucart, M{\"o}sta, Ramirez, Wright, Darbha
  \& Kasen}{Foucart et~al.}{2021}]{Foucart2021a}
Foucart F.,  M{\"o}sta P.,  Ramirez T.,  Wright A.~J.,  Darbha S.,   Kasen D.,
  2021, \mn@doi [Phys. Rev. D] {10.1103/physrevd.104.123010}, 104, 123010

\bibitem[\protect\citeauthoryear{Fujibayashi, Kiuchi, Nishimura, Sekiguchi  \&
  Shibata}{Fujibayashi et~al.}{2018}]{Fujibayashi2018a}
Fujibayashi S.,  Kiuchi K.,  Nishimura N.,  Sekiguchi Y.,   Shibata M.,  2018,
  \mn@doi [ApJ] {10.3847/1538-4357/aabafd}, 860, 64

\bibitem[\protect\citeauthoryear{Fujibayashi, Wanajo, Kiuchi, Kyutoku,
  Sekiguchi  \& Shibata}{Fujibayashi et~al.}{2020}]{Fujibayashi2020a}
Fujibayashi S.,  Wanajo S.,  Kiuchi K.,  Kyutoku K.,  Sekiguchi Y.,   Shibata
  M.,  2020, \mn@doi [ApJ] {10.3847/1538-4357/abafc2}, 901, 122

\bibitem[\protect\citeauthoryear{Fujibayashi, Kiuchi, Wanajo, Kyutoku,
  Sekiguchi  \& Shibata}{Fujibayashi et~al.}{2023}]{Fujibayashi2023a}
Fujibayashi S.,  Kiuchi K.,  Wanajo S.,  Kyutoku K.,  Sekiguchi Y.,   Shibata
  M.,  2023, \mn@doi [ApJ] {10.3847/1538-4357/ac9ce0}, 942, 39

\bibitem[\protect\citeauthoryear{Galeazzi, Kastaun, Rezzolla  \& Font}{Galeazzi
  et~al.}{2013}]{Galeazzi2013a}
Galeazzi F.,  Kastaun W.,  Rezzolla L.,   Font J.~A.,  2013, \mn@doi [Phys.
  Rev. D] {10.1103/PhysRevD.88.064009}, 88, 064009

\bibitem[\protect\citeauthoryear{Goodale, Allen, Lanfermann, Mass{\'o}, Radke,
  Seidel  \& Shalf}{Goodale et~al.}{2003}]{Goodale2003a}
Goodale T.,  Allen G.,  Lanfermann G.,  Mass{\'o} J.,  Radke T.,  Seidel E.,
  Shalf J.,  2003, in , Vol.~2565, High {{Performance Computing}} for
  {{Computational Science}} \textemdash{} {{VECPAR}} 2002.
{Springer Berlin Heidelberg}, pp 197--227

\bibitem[\protect\citeauthoryear{Gottlieb \& Shu}{Gottlieb \&
  Shu}{1998}]{Gottlieb1998a}
Gottlieb S.,  Shu C.-W.,  1998, \mn@doi [Math. Comput.]
  {10.1090/s0025-5718-98-00913-2}, 67, 73

\bibitem[\protect\citeauthoryear{Gourgoulhon, Grandcl{\'e}ment, Taniguchi,
  Marck  \& Bonazzola}{Gourgoulhon et~al.}{2001}]{Gourgoulhon2001a}
Gourgoulhon E.,  Grandcl{\'e}ment P.,  Taniguchi K.,  Marck J.-A.,   Bonazzola
  S.,  2001, \mn@doi [Phys. Rev. D] {10.1103/PhysRevD.63.064029}, 63, 064029

\bibitem[\protect\citeauthoryear{Haas et~al.,}{Haas et~al.}{2020}]{Haas2020a}
Haas R.,  et~al., 2020, The {{Einstein Toolkit}}, \url
  {https://zenodo.org/record/4298887}

\bibitem[\protect\citeauthoryear{Harten, Lax, {van Leer}  \& van Leer}{Harten
  et~al.}{1983}]{Harten1983a}
Harten A.,  Lax P.~D.,  {van Leer} B.,   van Leer B.,  1983, \mn@doi [SIAM
  Rev.] {10.1137/1025002}, 25, 35

\bibitem[\protect\citeauthoryear{Hayashi, Fujibayashi, Kiuchi, Kyutoku,
  Sekiguchi  \& Shibata}{Hayashi et~al.}{2022}]{Hayashi2022a}
Hayashi K.,  Fujibayashi S.,  Kiuchi K.,  Kyutoku K.,  Sekiguchi Y.,   Shibata
  M.,  2022, \mn@doi [Phys. Rev. D] {10.1103/PhysRevD.106.023008}, 106, 023008

\bibitem[\protect\citeauthoryear{Hebeler, Bogner, Furnstahl, Nogga  \&
  Schwenk}{Hebeler et~al.}{2011}]{Hebeler2011a}
Hebeler K.,  Bogner S.~K.,  Furnstahl R.~J.,  Nogga A.,   Schwenk A.,  2011,
  \mn@doi [Phys. Rev. C] {10.1103/PhysRevC.83.031301}, 83, 031301

\bibitem[\protect\citeauthoryear{Hebeler, Lattimer, Pethick  \&
  Schwenk}{Hebeler et~al.}{2013}]{Hebeler2013a}
Hebeler K.,  Lattimer J.~M.,  Pethick C.~J.,   Schwenk A.,  2013, \mn@doi [ApJ]
  {10.1088/0004-637X/773/1/11}, 773, 11

\bibitem[\protect\citeauthoryear{Henkel, Foucart, Raaijmakers  \&
  Nissanke}{Henkel et~al.}{2023}]{Henkel2023a}
Henkel A.,  Foucart F.,  Raaijmakers G.,   Nissanke S.,  2023, \mn@doi [Phys.
  Rev. D] {10.1103/PhysRevD.107.063028}, 107, 063028

\bibitem[\protect\citeauthoryear{Hinder et~al.,}{Hinder
  et~al.}{2013}]{Hinder2013a}
Hinder I.,  et~al., 2013, \mn@doi [Classical Quantum Gravity]
  {10.1088/0264-9381/31/2/025012}, 31, 025012

\bibitem[\protect\citeauthoryear{Hotokezaka, Kiuchi, Kyutoku, Okawa, Sekiguchi,
  Shibata  \& Taniguchi}{Hotokezaka et~al.}{2013a}]{Hotokezaka2013a}
Hotokezaka K.,  Kiuchi K.,  Kyutoku K.,  Okawa H.,  Sekiguchi Y.-i.,  Shibata
  M.,   Taniguchi K.,  2013a, \mn@doi [Phys. Rev. D]
  {10.1103/PhysRevD.87.024001}, 87, 024001

\bibitem[\protect\citeauthoryear{Hotokezaka, Kiuchi, Kyutoku, Muranushi,
  Sekiguchi, Shibata  \& Taniguchi}{Hotokezaka et~al.}{2013b}]{Hotokezaka2013b}
Hotokezaka K.,  Kiuchi K.,  Kyutoku K.,  Muranushi T.,  Sekiguchi Y.~I.,
  Shibata M.,   Taniguchi K.,  2013b, \mn@doi [Phys. Rev. D]
  {10.1103/PhysRevD.88.044026}, 88, 044026

\bibitem[\protect\citeauthoryear{Hu, Adams  \& Shu}{Hu et~al.}{2013}]{Hu2013a}
Hu X.~Y.,  Adams N.~A.,   Shu C.~W.,  2013, \mn@doi [J. Comput. Phys.]
  {10.1016/j.jcp.2013.01.024}, 242, 169

\bibitem[\protect\citeauthoryear{Hu et~al.,}{Hu et~al.}{2017}]{Hu2017a}
Hu L.,  et~al., 2017, \mn@doi [Sci. Bull.] {10.1016/J.SCIB.2017.10.006}, 62,
  1433

\bibitem[\protect\citeauthoryear{Huth, Wellenhofer  \& Schwenk}{Huth
  et~al.}{2021}]{Huth2021a}
Huth S.,  Wellenhofer C.,   Schwenk A.,  2021, \mn@doi [Phys. Rev. C]
  {10.1103/PhysRevC.103.025803}, 103, 025803

\bibitem[\protect\citeauthoryear{Huth et~al.,}{Huth et~al.}{2022}]{Huth2022a}
Huth S.,  et~al., 2022, \mn@doi [Nature] {10.1038/s41586-022-04750-w}, 606, 276

\bibitem[\protect\citeauthoryear{Just, Bauswein, Ardevol~Pulpillo, Goriely,
  Janka, Pulpillo, Goriely  \& Janka}{Just et~al.}{2015}]{Just2015a}
Just O.,  Bauswein A.,  Ardevol~Pulpillo R.,  Goriely S.,  Janka H.~T.,
  Pulpillo R.~A.,  Goriely S.,   Janka H.~T.,  2015, \mn@doi [MNRAS]
  {10.1093/mnras/stv009}, 448, 541

\bibitem[\protect\citeauthoryear{Kasen, Badnell  \& Barnes}{Kasen
  et~al.}{2013}]{Kasen2013a}
Kasen D.,  Badnell N.~R.,   Barnes J.,  2013, \mn@doi [ApJ]
  {10.1088/0004-637X/774/1/25}, 774, 25

\bibitem[\protect\citeauthoryear{Kasliwal et~al.,}{Kasliwal
  et~al.}{2017}]{Kasliwal2017a}
Kasliwal M.~M.,  et~al., 2017, \mn@doi [Science] {10.1126/SCIENCE.AAP9455},
  358, 1559

\bibitem[\protect\citeauthoryear{Kastaun \& Galeazzi}{Kastaun \&
  Galeazzi}{2015}]{Kastaun2015a}
Kastaun W.,  Galeazzi F.,  2015, \mn@doi [Phys. Rev. D]
  {10.1103/PhysRevD.91.064027}, 91, 064027

\bibitem[\protect\citeauthoryear{Kawaguchi, Shibata  \& Tanaka}{Kawaguchi
  et~al.}{2018}]{Kawaguchi2018a}
Kawaguchi K.,  Shibata M.,   Tanaka M.,  2018, \mn@doi [ApJ]
  {10.3847/2041-8213/AADE02}, 865, L21

\bibitem[\protect\citeauthoryear{Keller, Wellenhofer, Hebeler  \&
  Schwenk}{Keller et~al.}{2021}]{Keller2021a}
Keller J.,  Wellenhofer C.,  Hebeler K.,   Schwenk A.,  2021, \mn@doi [Phys.
  Rev. C] {10.1103/PhysRevC.103.055806}, 103, 055806

\bibitem[\protect\citeauthoryear{Keller, Hebeler  \& Schwenk}{Keller
  et~al.}{2023}]{Keller2023a}
Keller J.,  Hebeler K.,   Schwenk A.,  2023, \mn@doi [Phys. Rev. Lett.]
  {10.1103/PhysRevLett.130.072701}, 130, 072701

\bibitem[\protect\citeauthoryear{Kiuchi, Kawaguchi, Kyutoku, Sekiguchi  \&
  Shibata}{Kiuchi et~al.}{2020}]{Kiuchi2020a}
Kiuchi K.,  Kawaguchi K.,  Kyutoku K.,  Sekiguchi Y.,   Shibata M.,  2020,
  \mn@doi [Phys. Rev. D] {10.1103/PhysRevD.101.084006}, 101, 084006

\bibitem[\protect\citeauthoryear{Korobkin, Rosswog, Arcones  \&
  Winteler}{Korobkin et~al.}{2012}]{Korobkin2012a}
Korobkin O.,  Rosswog S.,  Arcones A.,   Winteler C.,  2012, \mn@doi [MNRAS]
  {10.1111/j.1365-2966.2012.21859.x}, 426, 1940

\bibitem[\protect\citeauthoryear{Kr{\"u}ger \& Foucart}{Kr{\"u}ger \&
  Foucart}{2020}]{Kruger2020a}
Kr{\"u}ger C.~J.,  Foucart F.,  2020, \mn@doi [Phys. Rev. D]
  {10.1103/PhysRevD.101.103002}, 101, 103002

\bibitem[\protect\citeauthoryear{Kulkarni}{Kulkarni}{2005}]{Kulkarni2005a}
Kulkarni S.~R.,  2005, \mn@doi [arXiv e-prints]
  {10.48550/arXiv.astro-ph/0510256}, astro-ph/0510256

\bibitem[\protect\citeauthoryear{Lattimer \& Lim}{Lattimer \&
  Lim}{2013}]{Lattimer2013a}
Lattimer J.~M.,  Lim Y.,  2013, \mn@doi [ApJ] {10.1088/0004-637X/771/1/51},
  771, 51

\bibitem[\protect\citeauthoryear{Lattimer \& Prakash}{Lattimer \&
  Prakash}{2016}]{Lattimer2016a}
Lattimer J.~M.,  Prakash M.,  2016, \mn@doi [Phys. Rep.]
  {10.1016/J.PHYSREP.2015.12.005}, 621, 127

\bibitem[\protect\citeauthoryear{Lattimer \& Schramm}{Lattimer \&
  Schramm}{1974}]{Lattimer1974a}
Lattimer J.~M.,  Schramm D.~N.,  1974, \mn@doi [ApJ] {10.1086/181612}, 192,
  L145

\bibitem[\protect\citeauthoryear{Lattimer \& Swesty}{Lattimer \&
  Swesty}{1991}]{Lattimer1991a}
Lattimer J.~M.,  Swesty D.~F.,  1991, \mn@doi [Nucl. Phys. A]
  {10.1016/0375-9474(91)90452-C}, 535, 331

\bibitem[\protect\citeauthoryear{Le~F{\`e}vre, Leifels, Reisdorf, Aichelin  \&
  Hartnack}{Le~F{\`e}vre et~al.}{2016}]{LeFevre2016a}
Le~F{\`e}vre A.,  Leifels Y.,  Reisdorf W.,  Aichelin J.,   Hartnack {\relax
  Ch}.,  2016, \mn@doi [Nucl. Phys. A] {10.1016/j.nuclphysa.2015.09.015}, 945,
  112

\bibitem[\protect\citeauthoryear{Legred, Chatziioannou, Essick, Han  \&
  Landry}{Legred et~al.}{2021}]{Legred2021a}
Legred I.,  Chatziioannou K.,  Essick R.,  Han S.,   Landry P.,  2021, \mn@doi
  [Phys. Rev. D] {10.1103/PhysRevD.104.063003}, 104, 063003

\bibitem[\protect\citeauthoryear{Lehner, Liebling, Palenzuela  \& Motl}{Lehner
  et~al.}{2016}]{Lehner2016a}
Lehner L.,  Liebling S.~L.,  Palenzuela C.,   Motl P.~M.,  2016, \mn@doi [Phys.
  Rev. D] {10.1103/PhysRevD.94.043003}, 94, 043003

\bibitem[\protect\citeauthoryear{Li \& Paczy{\'n}ski}{Li \&
  Paczy{\'n}ski}{1998}]{Li1998a}
Li L.-X.,  Paczy{\'n}ski B.,  1998, \mn@doi [ApJ] {10.1086/311680}, 507, L59

\bibitem[\protect\citeauthoryear{Lipunov et~al.,}{Lipunov
  et~al.}{2017}]{Lipunov2017a}
Lipunov V.~M.,  et~al., 2017, \mn@doi [ApJ] {10.3847/2041-8213/AA92C0}, 850, L1

\bibitem[\protect\citeauthoryear{L{\"o}ffler et~al.,}{L{\"o}ffler
  et~al.}{2012}]{Loffler2012a}
L{\"o}ffler F.,  et~al., 2012, \mn@doi [Classical Quantum Gravity]
  {10.1088/0264-9381/29/11/115001}, 29, 115001

\bibitem[\protect\citeauthoryear{Lynn, Tews, Carlson, Gandolfi, Gezerlis,
  Schmidt  \& Schwenk}{Lynn et~al.}{2016}]{Lynn2016a}
Lynn J.~E.,  Tews I.,  Carlson J.,  Gandolfi S.,  Gezerlis A.,  Schmidt K.~E.,
   Schwenk A.,  2016, \mn@doi [Phys. Rev. Lett.]
  {10.1103/PhysRevLett.116.062501}, 116, 062501

\bibitem[\protect\citeauthoryear{Maione et~al.,}{Maione
  et~al.}{2017}]{Maione2017a}
Maione F.,  et~al., 2017, \mn@doi [Phys. Rev. D] {10.1103/PhysRevD.96.063011},
  96, 063011

\bibitem[\protect\citeauthoryear{Margalit \& Metzger}{Margalit \&
  Metzger}{2017}]{Margalit2017a}
Margalit B.,  Metzger B.~D.,  2017, \mn@doi [ApJ] {10.3847/2041-8213/AA991C},
  850, L19

\bibitem[\protect\citeauthoryear{Metzger et~al.,}{Metzger
  et~al.}{2010}]{Metzger2010a}
Metzger B.~D.,  et~al., 2010, \mn@doi [MNRAS]
  {10.1111/j.1365-2966.2010.16864.x}, 406, 2650

\bibitem[\protect\citeauthoryear{Miller et~al.,}{Miller
  et~al.}{2019}]{Miller2019a}
Miller J.~M.,  et~al., 2019, \mn@doi [Phys. Rev. D]
  {10.1103/PHYSREVD.100.023008}, 100, 023008

\bibitem[\protect\citeauthoryear{Miller et~al.,}{Miller
  et~al.}{2021}]{Miller2021a}
Miller M.~C.,  et~al., 2021, \mn@doi [ApJ] {10.3847/2041-8213/ac089b}, 918, L28

\bibitem[\protect\citeauthoryear{Molero et~al.,}{Molero
  et~al.}{2021}]{Molero2021a}
Molero M.,  et~al., 2021, \mn@doi [MNRAS] {10.1093/MNRAS/STAB1429}, 505, 2913

\bibitem[\protect\citeauthoryear{Most \& Raithel}{Most \&
  Raithel}{2021}]{Most2021a}
Most E.~R.,  Raithel C.~A.,  2021, \mn@doi [Phys. Rev. D]
  {10.1103/PhysRevD.104.124012}, 104, 124012

\bibitem[\protect\citeauthoryear{Most, Papenfort, Weih  \& Rezzolla}{Most
  et~al.}{2020}]{Most2020a}
Most E.~R.,  Papenfort L.~J.,  Weih L.~R.,   Rezzolla L.,  2020, \mn@doi
  [MNRAS] {10.1093/mnrasl/slaa168}, 499, L82

\bibitem[\protect\citeauthoryear{Nathanail, Most  \& Rezzolla}{Nathanail
  et~al.}{2021}]{Nathanail2021a}
Nathanail A.,  Most E.~R.,   Rezzolla L.,  2021, \mn@doi [ApJ]
  {10.3847/2041-8213/ABDFC6}, 908, L28

\bibitem[\protect\citeauthoryear{Nedora, Bernuzzi, Radice, Perego, Endrizzi  \&
  Ortiz}{Nedora et~al.}{2019}]{Nedora2019a}
Nedora V.,  Bernuzzi S.,  Radice D.,  Perego A.,  Endrizzi A.,   Ortiz N.,
  2019, \mn@doi [ApJ] {10.3847/2041-8213/ab5794}, 886, L30

\bibitem[\protect\citeauthoryear{Nedora et~al.,}{Nedora
  et~al.}{2021a}]{Nedora2021c}
Nedora V.,  et~al., 2021a, \mn@doi [Classical Quantum Gravity]
  {10.1088/1361-6382/ac35a8}, 39, 015008

\bibitem[\protect\citeauthoryear{Nedora, Radice, Bernuzzi, Perego, Daszuta,
  Endrizzi, Prakash  \& Schianchi}{Nedora et~al.}{2021b}]{Nedora2021a}
Nedora V.,  Radice D.,  Bernuzzi S.,  Perego A.,  Daszuta B.,  Endrizzi A.,
  Prakash A.,   Schianchi F.,  2021b, \mn@doi [MNRAS] {10.1093/mnras/stab2004},
  506, 5908

\bibitem[\protect\citeauthoryear{Nedora et~al.,}{Nedora
  et~al.}{2021c}]{Nedora2021b}
Nedora V.,  et~al., 2021c, \mn@doi [ApJ] {10.3847/1538-4357/abc9be}, 906, 98

\bibitem[\protect\citeauthoryear{Neilsen, Liebling, Anderson, Lehner, O'Connor
  \& Palenzuela}{Neilsen et~al.}{2014}]{Neilsen2014a}
Neilsen D.,  Liebling S.~L.,  Anderson M.,  Lehner L.,  O'Connor E.,
  Palenzuela C.,  2014, \mn@doi [Phys. Rev. D] {10.1103/PhysRevD.89.104029},
  89, 104029

\bibitem[\protect\citeauthoryear{Newman \& Penrose}{Newman \&
  Penrose}{1962}]{Newman1962a}
Newman E.,  Penrose R.,  1962, \mn@doi [J. Math. Phys.] {10.1063/1.1724257}, 3,
  566

\bibitem[\protect\citeauthoryear{Paschalidis, East, Pretorius  \&
  Shapiro}{Paschalidis et~al.}{2015}]{Paschalidis2015a}
Paschalidis V.,  East W.~E.,  Pretorius F.,   Shapiro S.~L.,  2015, \mn@doi
  [Phys. Rev. D] {10.1103/PhysRevD.92.121502}, 92, 121502

\bibitem[\protect\citeauthoryear{Perego, Rosswog, Cabez{\'o}n, Korobkin,
  K{\"a}ppeli, Arcones  \& Liebend{\"o}rfer}{Perego et~al.}{2014}]{Perego2014a}
Perego A.,  Rosswog S.,  Cabez{\'o}n R.~M.,  Korobkin O.,  K{\"a}ppeli R.,
  Arcones A.,   Liebend{\"o}rfer M.,  2014, \mn@doi [MNRAS]
  {10.1093/mnras/stu1352}, 443, 3134

\bibitem[\protect\citeauthoryear{Pian et~al.,}{Pian et~al.}{2017}]{Pian2017a}
Pian E.,  et~al., 2017, \mn@doi [Nature] {10.1038/NATURE24298}, 551, 67

\bibitem[\protect\citeauthoryear{Pozanenko et~al.,}{Pozanenko
  et~al.}{2018}]{Pozanenko2018a}
Pozanenko A.~S.,  et~al., 2018, \mn@doi [ApJ] {10.3847/2041-8213/aaa2f6}, 852,
  L30

\bibitem[\protect\citeauthoryear{Raaijmakers et~al.,}{Raaijmakers
  et~al.}{2019}]{Raaijmakers2019a}
Raaijmakers G.,  et~al., 2019, \mn@doi [ApJ] {10.3847/2041-8213/ab451a}, 887,
  L22

\bibitem[\protect\citeauthoryear{Raaijmakers et~al.,}{Raaijmakers
  et~al.}{2021}]{Raaijmakers2021a}
Raaijmakers G.,  et~al., 2021, \mn@doi [ApJ] {10.3847/2041-8213/ac089a}, 918,
  L29

\bibitem[\protect\citeauthoryear{Radice \& Rezzolla}{Radice \&
  Rezzolla}{2012}]{Radice2012a}
Radice D.,  Rezzolla L.,  2012, \mn@doi [A\&A] {10.1051/0004-6361/201219735},
  547, A26

\bibitem[\protect\citeauthoryear{Radice, Rezzolla  \& Galeazzi}{Radice
  et~al.}{2014a}]{Radice2014a}
Radice D.,  Rezzolla L.,   Galeazzi F.,  2014a, \mn@doi [Classical Quantum
  Gravity] {10.1088/0264-9381/31/7/075012}, 31, 075012

\bibitem[\protect\citeauthoryear{Radice, Rezzolla  \& Galeazzi}{Radice
  et~al.}{2014b}]{Radice2014b}
Radice D.,  Rezzolla L.,   Galeazzi F.,  2014b, \mn@doi [MNRAS]
  {10.1093/mnrasl/slt137}, 437, L46

\bibitem[\protect\citeauthoryear{Radice, Bernuzzi  \& Ott}{Radice
  et~al.}{2016}]{Radice2016a}
Radice D.,  Bernuzzi S.,   Ott C.~D.,  2016, \mn@doi [Phys. Rev. D]
  {10.1103/PhysRevD.94.064011}, 94, 064011

\bibitem[\protect\citeauthoryear{Radice, Perego, Hotokezaka, Fromm, Bernuzzi
  \& Roberts}{Radice et~al.}{2018}]{Radice2018a}
Radice D.,  Perego A.,  Hotokezaka K.,  Fromm S.~A.,  Bernuzzi S.,   Roberts
  L.~F.,  2018, \mn@doi [ApJ] {10.3847/1538-4357/aaf054}, 869, 130

\bibitem[\protect\citeauthoryear{Radice, Bernuzzi  \& Perego}{Radice
  et~al.}{2020}]{Radice2020a}
Radice D.,  Bernuzzi S.,   Perego A.,  2020, \mn@doi [Annu. Rev. Nucl. Part.
  Sci.] {10.1146/annurev-nucl-013120-114541}, 70, 95

\bibitem[\protect\citeauthoryear{Raithel, Paschalidis  \& {\"O}zel}{Raithel
  et~al.}{2021}]{Raithel2021a}
Raithel C.~A.,  Paschalidis V.,   {\"O}zel F.,  2021, \mn@doi [Phys. Rev. D]
  {10.1103/PhysRevD.104.063016}, 104, 063016

\bibitem[\protect\citeauthoryear{Reisswig \& Pollney}{Reisswig \&
  Pollney}{2011}]{Reisswig2011a}
Reisswig C.,  Pollney D.,  2011, \mn@doi [Classical Quantum Gravity]
  {10.1088/0264-9381/28/19/195015}, 28, 195015

\bibitem[\protect\citeauthoryear{Rezzolla \& Takami}{Rezzolla \&
  Takami}{2016}]{Rezzolla2016a}
Rezzolla L.,  Takami K.,  2016, \mn@doi [Phys. Rev. D]
  {10.1103/PhysRevD.93.124051}, 93, 124051

\bibitem[\protect\citeauthoryear{Rezzolla, Most  \& Weih}{Rezzolla
  et~al.}{2018}]{Rezzolla2018a}
Rezzolla L.,  Most E.~R.,   Weih L.~R.,  2018, \mn@doi [ApJ]
  {10.3847/2041-8213/AAA401}, 852, L25

\bibitem[\protect\citeauthoryear{Riley et~al.,}{Riley
  et~al.}{2019}]{Riley2019a}
Riley T.~E.,  et~al., 2019, \mn@doi [ApJ] {10.3847/2041-8213/ab481c}, 887, L21

\bibitem[\protect\citeauthoryear{Riley et~al.,}{Riley
  et~al.}{2021}]{Riley2021a}
Riley T.~E.,  et~al., 2021, \mn@doi [ApJ] {10.3847/2041-8213/ac0a81}, 918, L27

\bibitem[\protect\citeauthoryear{{Roca-Maza}, Vi{\~n}as, Centelles, Agrawal,
  Col{\`o}, Paar, Piekarewicz  \& Vretenar}{{Roca-Maza}
  et~al.}{2015}]{Roca-Maza2015a}
{Roca-Maza} X.,  Vi{\~n}as X.,  Centelles M.,  Agrawal B.~K.,  Col{\`o} G.,
  Paar N.,  Piekarewicz J.,   Vretenar D.,  2015, \mn@doi [Phys. Rev. C]
  {10.1103/PhysRevC.92.064304}, 92, 064304

\bibitem[\protect\citeauthoryear{Rosswog}{Rosswog}{2013}]{Rosswog2013a}
Rosswog S.,  2013, \mn@doi [Philos. Transactions Royal Soc. A]
  {10.1098/rsta.2012.0272}, 371, 20120272

\bibitem[\protect\citeauthoryear{Rosswog, Liebendoerfer, Thielemann, Davies,
  Benz  \& Piran}{Rosswog et~al.}{1999}]{Rosswog1999a}
Rosswog S.,  Liebendoerfer M.,  Thielemann F.~K.,  Davies M.~B.,  Benz W.,
  Piran T.,  1999, \mn@doi [A\&A] {10.48550/arXiv.astro-ph/9811367}, 341, 499

\bibitem[\protect\citeauthoryear{Rosswog, Sollerman, Feindt, Goobar, Korobkin,
  Wollaeger, Fremling  \& Kasliwal}{Rosswog et~al.}{2018}]{Rosswog2018a}
Rosswog S.,  Sollerman J.,  Feindt U.,  Goobar A.,  Korobkin O.,  Wollaeger R.,
   Fremling C.,   Kasliwal M.~M.,  2018, \mn@doi [A\&A]
  {10.1051/0004-6361/201732117}, 615

\bibitem[\protect\citeauthoryear{Russotto et~al.,}{Russotto
  et~al.}{2016}]{Russotto2016a}
Russotto P.,  et~al., 2016, \mn@doi [Phys. Rev. C]
  {10.1103/PhysRevC.94.034608}, 94, 034608

\bibitem[\protect\citeauthoryear{Schneider, Roberts  \& Ott}{Schneider
  et~al.}{2017}]{Schneider2017a}
Schneider A.~S.,  Roberts L.~F.,   Ott C.~D.,  2017, \mn@doi [Phys. Rev. C]
  {10.1103/PhysRevC.96.065802}, 96, 065802

\bibitem[\protect\citeauthoryear{Schneider, Roberts  \& Ott}{Schneider
  et~al.}{2018}]{Schneider2018a}
Schneider A. d.~S.,  Roberts L.~F.,   Ott C.~D.,  2018, {{SROEOS}}, \url
  {https://bitbucket.org/andschn/sroeos/src/master/}

\bibitem[\protect\citeauthoryear{Schneider, Roberts, Ott  \&
  O'Connor}{Schneider et~al.}{2019}]{Schneider2019a}
Schneider A.~S.,  Roberts L.~F.,  Ott C.~D.,   O'Connor E.,  2019, \mn@doi
  [Phys. Rev. C] {10.1103/PhysRevC.100.055802}, 100, 055802

\bibitem[\protect\citeauthoryear{Schnetter, Hawley  \& Hawke}{Schnetter
  et~al.}{2004}]{Schnetter2004a}
Schnetter E.,  Hawley S.~H.,   Hawke I.,  2004, \mn@doi [Classical Quantum
  Gravity] {10.1088/0264-9381/21/6/014}, 21, 1465

\bibitem[\protect\citeauthoryear{Sekiguchi, Kiuchi, Kyutoku  \&
  Shibata}{Sekiguchi et~al.}{2015}]{Sekiguchi2015a}
Sekiguchi Y.,  Kiuchi K.,  Kyutoku K.,   Shibata M.,  2015, \mn@doi [Phys. Rev.
  D] {10.1103/PhysRevD.91.064059}, 91, 064059

\bibitem[\protect\citeauthoryear{Sekiguchi, Kiuchi, Kyutoku, Shibata  \&
  Taniguchi}{Sekiguchi et~al.}{2016}]{Sekiguchi2016a}
Sekiguchi Y.,  Kiuchi K.,  Kyutoku K.,  Shibata M.,   Taniguchi K.,  2016,
  \mn@doi [Phys. Rev. D] {10.1103/PhysRevD.93.124046}, 93, 124046

\bibitem[\protect\citeauthoryear{Shen, Toki, Oyamatsu  \& Sumiyoshi}{Shen
  et~al.}{1998}]{Shen1998a}
Shen H.,  Toki H.,  Oyamatsu K.,   Sumiyoshi K.,  1998, \mn@doi [Nucl. Phys. A]
  {10.1016/S0375-9474(98)00236-X}, 637, 435

\bibitem[\protect\citeauthoryear{Shibata \& Nakamura}{Shibata \&
  Nakamura}{1995}]{Shibata1995a}
Shibata M.,  Nakamura T.,  1995, \mn@doi [Phys. Rev. D]
  {10.1103/PhysRevD.52.5428}, 52, 5428

\bibitem[\protect\citeauthoryear{Shibata, Zhou, Kiuchi  \& Fujibayashi}{Shibata
  et~al.}{2019}]{Shibata2019a}
Shibata M.,  Zhou E.,  Kiuchi K.,   Fujibayashi S.,  2019, \mn@doi [Phys. Rev.
  D] {10.1103/PhysRevD.100.023015}, 100, 023015

\bibitem[\protect\citeauthoryear{Smartt et~al.,}{Smartt
  et~al.}{2017}]{Smartt2017a}
Smartt S.~J.,  et~al., 2017, \mn@doi [Nature] {10.1038/NATURE24303}, 551, 75

\bibitem[\protect\citeauthoryear{Somasundaram, Drischler, Tews  \&
  Margueron}{Somasundaram et~al.}{2021}]{Somasundaram2021a}
Somasundaram R.,  Drischler C.,  Tews I.,   Margueron J.,  2021, \mn@doi [Phys.
  Rev. C] {10.1103/PhysRevC.103.045803}, 103, 045803

\bibitem[\protect\citeauthoryear{Stergioulas, Bauswein, Zagkouris  \&
  Janka}{Stergioulas et~al.}{2011}]{Stergioulas2011a}
Stergioulas N.,  Bauswein A.,  Zagkouris K.,   Janka H.~T.,  2011, \mn@doi
  [MNRAS] {10.1111/j.1365-2966.2011.19493.x}, 418, 427

\bibitem[\protect\citeauthoryear{Suresh \& Huynh}{Suresh \&
  Huynh}{1997}]{Suresh1997a}
Suresh A.,  Huynh H.~T.,  1997, \mn@doi [J. Comput. Phys.]
  {10.1006/jcph.1997.5745}, 136, 83

\bibitem[\protect\citeauthoryear{Takami, Rezzolla  \& Baiotti}{Takami
  et~al.}{2014}]{Takami2014a}
Takami K.,  Rezzolla L.,   Baiotti L.,  2014, \mn@doi [Phys. Rev. Lett.]
  {10.1103/PhysRevLett.113.091104}, 113, 091104

\bibitem[\protect\citeauthoryear{Takami, Rezzolla  \& Baiotti}{Takami
  et~al.}{2015}]{Takami2015a}
Takami K.,  Rezzolla L.,   Baiotti L.,  2015, \mn@doi [Phys. Rev. D]
  {10.1103/PhysRevD.91.064001}, 91, 064001

\bibitem[\protect\citeauthoryear{Tanaka \& Hotokezaka}{Tanaka \&
  Hotokezaka}{2013}]{Tanaka2013a}
Tanaka M.,  Hotokezaka K.,  2013, \mn@doi [ApJ] {10.1088/0004-637X/775/2/113},
  775, 113

\bibitem[\protect\citeauthoryear{Tanvir et~al.,}{Tanvir
  et~al.}{2017}]{Tanvir2017a}
Tanvir N.~R.,  et~al., 2017, \mn@doi [ApJ] {10.3847/2041-8213/AA90B6}, 848, L27

\bibitem[\protect\citeauthoryear{Tews, Kr{\"u}ger, Hebeler  \& Schwenk}{Tews
  et~al.}{2013}]{Tews2013a}
Tews I.,  Kr{\"u}ger T.,  Hebeler K.,   Schwenk A.,  2013, \mn@doi [Phys. Rev.
  Lett.] {10.1103/PhysRevLett.110.032504}, 110, 032504

\bibitem[\protect\citeauthoryear{Troja et~al.,}{Troja
  et~al.}{2017}]{Troja2017a}
Troja E.,  et~al., 2017, \mn@doi [Nature] {10.1038/NATURE24290}, 551, 71

\bibitem[\protect\citeauthoryear{Tsang et~al.,}{Tsang
  et~al.}{2012}]{Tsang2012a}
Tsang M.~B.,  et~al., 2012, \mn@doi [Phys. Rev. C]
  {10.1103/PhysRevC.86.015803}, 86, 015803

\bibitem[\protect\citeauthoryear{Utsumi et~al.,}{Utsumi
  et~al.}{2017}]{Utsumi2017a}
Utsumi Y.,  et~al., 2017, \mn@doi [PASJ] {10.1093/PASJ/PSX118}, 69, 101

\bibitem[\protect\citeauthoryear{Valenti et~al.,}{Valenti
  et~al.}{2017}]{Valenti2017a}
Valenti S.,  et~al., 2017, \mn@doi [ApJ] {10.3847/2041-8213/AA8EDF}, 848, L24

\bibitem[\protect\citeauthoryear{Wellenhofer, Holt  \& Kaiser}{Wellenhofer
  et~al.}{2016}]{Wellenhofer2016a}
Wellenhofer C.,  Holt J.~W.,   Kaiser N.,  2016, \mn@doi [Phys. Rev. C]
  {10.1103/PhysRevC.93.055802}, 93, 055802

\bibitem[\protect\citeauthoryear{Yasin, Sch{\"a}fer, Arcones  \& Schwenk}{Yasin
  et~al.}{2020}]{Yasin2020a}
Yasin H.,  Sch{\"a}fer S.,  Arcones A.,   Schwenk A.,  2020, \mn@doi [Phys.
  Rev. Lett.] {10.1103/PhysRevLett.124.092701}, 124, 092701

\makeatother
\end{thebibliography}

\bsp
\label{lastpage}
\end{document}